\documentclass[%
 amsmath,amssymb,
 aps,
]{revtex4-2}
\usepackage{graphicx}
\usepackage{newtxtext}
\usepackage{newtxmath}
\usepackage{natbib}
\usepackage{hyperref}
\hypersetup{
    colorlinks = true,
    urlcolor   = blue,
    citecolor  = black,
}

\newcommand{\RomanNumeralCaps}[1]
\linenumbers

\usepackage{enumitem}
\usepackage{verbatim}

\usepackage{graphicx,color}  
\usepackage{dcolumn}   
\usepackage{bm}        
\usepackage{float}
\usepackage{cleveref}
\usepackage{subcaption}

\usepackage{mathtools}
\usepackage{graphicx}
\usepackage{natbib,color}

\usepackage{graphicx}
\usepackage{epstopdf, epsfig}
\usepackage{verbatim}
\usepackage{xfrac}
\usepackage{enumitem}
\usepackage{mathtools}
\usepackage{bm}
\usepackage{color}
\usepackage{mdframed}
\usepackage{booktabs}
\usepackage{wrapfig}
\usepackage[table,x11names]{xcolor}
\usepackage{nicefrac}
\usepackage{subcaption}
\usepackage{cases}
\usepackage{soul}
	\setstcolor{red}
\usepackage{siunitx}

\newcommand{\ci}{\mathrm{i}}

\newcommand{\p}{\partial}
\allowdisplaybreaks

\newcommand{\crefrangeconjunction}{--}

\newcommand{\pard}[2]{\frac{\partial #1}{\partial #2}}

\newcommand{\pardo}[2]{\frac{\partial^2 #1}{\partial #2^2}}

\newcommand{\Pint}{\mathop{\mathrlap{\pushP}}\!\int}
\newcommand{\pushP}{\mathchoice
  {\mkern1mu -}
  {\scriptstyle -}
  {\scriptscriptstyle -}
  {\scriptscriptstyle -}
}
\newcommand{\ccdot}{\text{\raise 0.2ex\hbox{\fontsize{3.5}{4}\selectfont $\bullet$}}}
\newcommand{\ccdote}{\text{\raise 0.3ex\hbox{\fontsize{3.5}{4}\selectfont $\bullet$}}}
\newcommand*\diff{\mathop{}\!\mathrm{d}}

\begin{document}
\preprint{APS}
\title{Hybrid Wave/Current Energy Harvesting with a Flexible Piezoelectric Plate}

\author{Kourosh Shoele}
\email{kshoele@eng.famu.fsu.edu}
\affiliation{Department of Mechanical Engineering, FAMU-FSU College of Engineering, Tallahassee, FL, 32310, USA}

\begin{abstract}
We investigate the dynamics and energy production capability of a  flexible piezoelectric plate submerged close to the free surface and exposed to incident head gravity waves and current. {
A theoretical model is derived in which the flag and its wake are represented with a vortex line while the body of the fluid is considered to be inviscid. The model is employed to describe the hydrodynamic interactions between a flexible plate, its wake, gravity incident waves and the current.} The model reveals two distinct vibration states of a piezoelectric device corresponding to almost similar optimal energy production levels. The first is associated with the cantilever fluttering mode of the plate with limited dependency on the plate's flexibility across different Froude numbers and incoming wave frequencies. The other resembles the flow-induced flapping mode in more flexible plates, with the energy output showing a higher dependency on plate flexibility. The concurrent existence of these two energetic modes allows adjustment of the plate length to consistently achieve the maximum energy production level across different flow conditions. The role of the Froude number of the system's responses is explored and correlated to the appearance of gravity wave groups on the surface, each propagating with a different wavenumber. It is shown that a submergence depth of less than half of the body length is required to reach a high energetic condition in subcritical and critical flows. Finally, the optimal inductive and resistive values are related to proper matching between flow, mechanical and electrical timescales. 
\end{abstract}

\maketitle
\section{Introduction}
The wave energy resource is significant and could supply 10 to 20\% of the world’s energy demand \citep{aderinto2018ocean}. Different technologies have been proposed to extract renewable energy from ocean current and ocean waves \citep{babarit2017ocean}. Still, few technologies can harvest energy from these two resources concurrently. One promising candidate is to use the coupling between fluid and structure in terms of vortex-induced vibration, galloping and fluttering response to either directly enable energy transfer from the incoming current \citep{zhu2009mode} or actively modify the impedance of the device to capture ocean energy more efficiently \citep{falnes2020ocean,ringwood2020wave}. Deployable and flexible structures are good candidates to enhance the performance of energy harvesting devices while addressing the reliability and survivability challenges facing conventional designs based on rigid-body systems \citep{pecher2017handbook}. The flexible wave energy converters like the wave carpet \citep{koola2003dynamics,alam2012nonlinear} and piezoelectric wave energy converter (PWEC) \citep{jbaily2015piezoelectric} use the cyclic wave action to generate an AC electrical energy. These devices are easily adjustable and can be employed as wave attenuation devices and hybrid wave/current energy harvesters on fixed and moving marine structures. They can also be combined with other devices to enhance the locomotion of marine vehicles \citep{collins2021flexible}. 

A flexible thin structure with distributed wave energy converters, analogous to a piezoelectric plate, is a canonical rendition of a wide range of the above multi-segmented flexible energy harvesting concepts  \citep{erturk2011piezoelectric}. Piezoelectric thin structures are employed to directly harvest wave energy \citep{mutsuda2019application,viet2017review}, or they are used as a model to study the asymptotic response of multi-segmented conventional devices, such as  M4 and Pelamis, wherein the energy is captured from differential deformation of adjacent segments  \citep{stansby2015capture,peng2020experimental}. These systems, in the limit of many interconnecting bodies, resemble the canonical piezoelectric plate with infinitesimal neighboring electrical circuits in which a simple mathematical model can capture the role of mechanical-to-electrical conversion as well as the resistance capacitance and inductance components of the electrical circuit \citep{michelin2013energy,doare2011piezoelectric}. Another related power harvesting system with similar principles is electro-active polymers \citep{babarit2013hydro}. All of these technologies have the advantage of directly transforming the mechanical actuation to electrical output and therefore eliminating the need for including complex mechanical power take-off and power conversion systems \citep{renzi2021niche}.   The coupled hydro-electro-mechanic response of these structures has been studied near the free surface \cite{renzi_hydroelectromechanical_2016}, next to vertical wall \citep{zheng2020wave} and as an array of energy harvesters \citep{mutsuda2013ocean}. Furthermore, Previous research has explored using flexible plates as adjustable floaters and wave energy dissipation devices \citep{Zheng2021,selvan2020wave}. {In particular, the use of floating plates for harvesting wave energy is explored in \cite{michele_wave_2020,michele2022waveb} wherein a  theoretical model based on structural mode shapes of a circular plate is proposed to examine the wave power extraction capacity of these concepts.}

In a recent study conducted by \cite{mougel2020flutter}, it was discovered that the flapping dynamics of a flexible plate in potential flow are significantly influenced by the presence of the free surface and resonant conditions between the plate and surface gravity waves. 
Moreover, previous research by \cite{renzi2016hydroelectromechanical} highlighted that optimal energy harvesting by a piezoelectric plate in a purely harmonic incoming wave relies on constructive matching between the plate's natural frequency and the wave frequency. Building upon these findings, the current study explores how the flow interaction among a flexible piezoelectric plate, its wake, and the free surface can be utilized to manipulate the dynamic characteristics of hybrid wave/current energy harvesters and enhance their energy-capturing efficiency.

A closely related problem is a hydrofoil oscillation near the free surface as a passive propulsor or energy harvesting device \citep{rozhdestvensky_aerohydrodynamics_2003}. In \cite{grue_propulsion_1988} and \cite{crimi1964forces}, a mathematical model was proposed for the propulsion of a moving and oscillating foil in close proximity to a free surface under two different scenarios: with incoming waves and without incoming waves. In the absence of incoming waves and a stationary free surface, the study revealed that the momentum transfer between the foil and the flow could result in either positive or negative values, leading to an increase or decrease in the thrust force. However, regardless of the specific scenario, the generation of waves always results in a significant amount of wasted energy. With the presence of surface waves, it was shown that up to 75\% of the incoming wave energy could be potentially captured and leveraged toward propulsion. Similarly, it has been explored that prescribed deforming fishes near the free surface can benefit from the surface waves to adjust their propulsion \citep{reece_swimming_1964,fish1999review, shoele2015drafting}. It is interesting to study how through proper selection of structural parameters, one can take advantage of the combined effects of current and wave for energy harvesting applications and achieve optimal operating conditions.  

{
This article employs a mathematical model of a flexible piezoelectric flexible plate placed at a finite depth below the free surface to study the energy harvesting regimes. A wide range of incoming wave frequency, submergence depth and current are investigated. The effects of mechanical and electrical parameters on the energy transfer between flow and structure will be quantified and major response modes will be examined.}  Finally, the conclusion and discussion about future directions will be presented.

\section{Problem Formulation}

\subsection{Configuration}
\begin{figure}
\centering {
\includegraphics[width=0.8\textwidth]{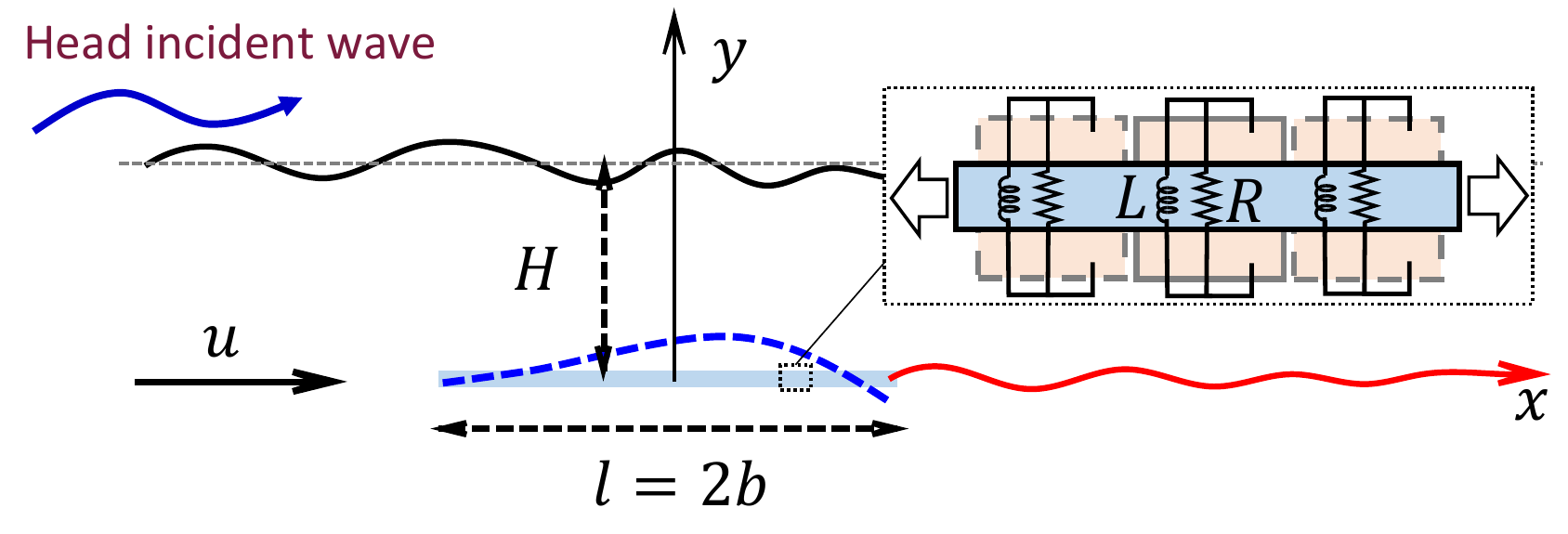}}
\caption{ Schematic configuration of a piezoelectric plate beneath the free surface in uniform flow exposed to incident harmonic wave. {In this study, the incident wave is assumed to propagate in the direction of the current, which is also referred to as the head wave \citep{grue1985wave}.}
\label{Schematic}  }
\end{figure}

We consider the dynamics of a two-dimensional thin {neutrally-buoyant} flexible plate of length $2b$ with $b$ being the half cord length placed in a uniform axial flow of velocity $u$ and density $\rho$ at a distance $H$ beneath the mean free surface (Fig. \ref{Schematic}). The flow is assumed to be incompressible, inviscid and infinitely deep. {The infinite depth assumption is made to simplify the theoretical modeling. In addition, the infinite depth assumption is appropriate for the current problem since, in the anticipated energy harvesting applications, the plate's length is much smaller than the water depth. }
Here,  we assume the clamped boundary condition at the leading edge and free boundary condition at the trailing edge of the plate. In addition, it is considered that both surfaces of the plate are covered with infinitesimal piezoelectric patches with segmentation lengths much smaller than $b$. Each segment is attached to an independent energy harvesting system with linear resistance ($R$), Capacitance ($C$) and inductance ($L$) elements. The sandwich plate's mass per unit length is $m_s$ and its equivalent bending stiffness is $k_b$. Since the plate is very thin, the inertial effect of the plate mass is negligible and therefore is not considered. 

\subsection{Governing equations }
{
It is assumed that the oscillation amplitudes 
and the amplitudes of the incoming waves are small. This allows us to use linearized equations of the plate and flow boundary conditions. We denote the vertical deflection of the plate with $Y \left(x,t\right)$ and the voltage between the top and the electric voltage difference between a small piezoelectric patched on the top and bottom surfaces of the flag with $V \left(x,t\right)$. The piezoelectric patch is connected to the output circuit with resistance and inductance elements, as shown in Figure \ref{Schematic}. {It is assumed that $Y \left(x,t\right) =O\left(\epsilon_p b\right)\, \ll b$, for a small $\epsilon_p$ value. Here, $\epsilon_p$ is related to  incoming wave amplitude as $\epsilon_p\,\approx {A_0}/{b}$  with $A_0$ being the incoming wave amplitude.} For a thin submerged plate, the structural inertial effects are negligible compared to the bending effect and fluid forces; therefore, the flag is assumed to be massless. Furthermore, it is assumed that the gravity and buoyancy forces cancel each other for the neutrally-buoyant flag considered in this study and the initial configuration of the plate is forced free.} {The non-dimensional variables shown with $\hat{\,\,\,\,}$ are defined as,

\begin{equation}
\hat{x}\,=\,\frac{x}{b}, \hspace{0.5cm} \hat{Y}\,=\,\frac{Y}{b}, \hspace{0.5cm} 
\hat{t}\,=\,\frac{u t}{b}, \hspace{.5cm} \hat{V}\,=\,\frac{V}{u\sqrt{\rho b/C}}.
\label{eq:nondim0}
\end{equation}

Here, the non-dimensionalization is done based on  $b$, $b/u$, $\rho$, $u\sqrt{\rho b/C}$ and $u \sqrt{\rho b C}$ as reference length, time, density, voltage and charge density. Henceforth, we adopt the same notation for the non-dimensional variables of $x$, $t$, $Y$ and $V$ as their dimensional counterparts for brevity.   The non-dimensional linearized electrical-mechanical equations of the piezoelectric plate can be written as \citep{xia2015fluid,shoele2016energy}, 
\begin{equation}
\frac{1}{{U^*}^2},\frac{\p^4 Y}{\p x^4} \, -\, \frac{\alpha}{U^*}\, \pardo{V}{x}\, =\, -[P], \label{eq:EOM} \end{equation}
\begin{equation}
\beta \frac{\p^2 V}{\p t^2} \, +\, \frac{\p V}{\p t}\,+\, \beta \tau^2 V \, +\,  \frac{\alpha\beta}{U^*} \, \frac{\p^4 Y}{\p x^2 \p t^2}\,=\,0,
\label{eq:EOE}
\end{equation}
where $U^*$ is the non-dimensional free-stream velocity, $\alpha$ is the coupling coefficient, $\beta$ quantifies the resistant property of the electrical circuit , and $\tau$ represents the inductance property of the electrical circuit \citep{thomas2009vibrations,xia2015fluid}. These characteristic parameters are defined as,
\begin{align}  
         U^*=u b \sqrt{\frac{\rho b}{k_b}}\,\,\,\,\,\,  \alpha=\frac{\chi}{\sqrt{k_b C}};\,\,\,\,\,\,    \beta=\frac{C\,u\,R}{b}; \,\,\,\,\,\,   \tau=\frac{b}{u\sqrt{C\,L}}, \label{eq:nondim}
  \end{align}   
where $\chi$ is the coupling-coefficient of piezoelectric patches {and $[P]=P^+-P^-$ is the pressure jump across the thin plate.} 
The {clamped} and free boundary conditions at the $x=-1$ and $x=1$ are defined as,
\begin{align}    
Y=\,0,\, \hspace{2.1cm} \frac{\p Y}{\p x}\, =\, 0,\, &\text{ at } x=-1, \label{eq:BCfixed}\\
\frac{1}{{U^*}} \frac{\p^2 Y}{\p x^2}-\alpha V=\, 0,\, \hspace{0.25cm} 
\frac{1}{{U^*}} \frac{\p^3 Y}{\p x^3}-\alpha \frac{\p V}{\p x}=\, 0,\, &\text{ at } x=\,\, \,\, 1. \label{eq:BCfree2}
\end{align}}


As a starting point in describing the flow, we follow the procedure proposed by \cite{crimi1964forces}. The flow is assumed to be potential and the effect of viscosity is confined to the thin two-dimensional vortex sheet along the plate and the deformable free vortex sheet associated with the wake of the plate. The former part is known as the bounded vortex sheet, and the latter is the free vortex sheet. This model is originally proposed to study the dynamic of thin airfoils and flexible plates \citep{thwaites1960incompressible,nitsche1994numerical,pullin2004unsteady,alben_optimal_2008}. Moreover, because the plate deflection is on the order of $O\left(\epsilon\right)$, the wake is approximately in the  $y = 0$ plane. As will be discussed shortly later, this model can be used to relate the average flow velocity in the fluid domain to the vortex sheet strength $\Lambda$ using Biot-Savart kernel \citep{saffman1995vortex}. Knowing the $\Lambda$ along the plate, the pressure jump across the plate, $[P]$, can be calculated from the unsteady Bernoulli equation as,
\begin{align}    
-\pard{[P]}{x}=\pard{\Lambda}{t} \,+\, \pard{\Lambda}{x} ,\,\,\,\,\,\,\,\,{\text{at}\,\,\, y=0. }
\label{eq:EOP}
\end{align}

A potential function $\Phi\left(x,y,t\right)$ is used to specify the perturbed flow caused by the presence of the plate, its wake and the incoming waves and current. The velocity potential satisfies the Laplace equation inside the flow region and linearized boundary conditions on the free surface \citep{newman2018marine,grue_propulsion_1988,haskind1954wave},
\begin{equation}    
\nabla^2 \Phi\left(x,y,t\right) =\, 0, \/ \label{eq:POT} 
\end{equation}
\begin{equation} 
\pard{\eta}{t} \,+ \frac{\p \eta}{\p x} =\frac{\p \Phi}{\p y},\,\,\,\,\,\,\,\,  \text{at}\,\,\, y=h   \label{eq:KBC}
\end{equation}
\begin{equation} 
\pard{\Phi}{t}\,+ \frac{\p \Phi}{\p x} =\, \frac{-1}{Fr^2}\eta,\,\,\,\,\,\,\,\,\text{at\/\/}\,\,\, y=h  \label{eq:DBC} 
\end{equation}
where $\eta(x,t)$ is the height of the free surface disturbance, $h=H/b$ is the normalized submergence depth and $Fr$ is the Froude number defined as,
\begin{align} 
Fr\,&=\,\frac{u}{\sqrt{gb}}\,=\,\frac{u}{\sqrt{g\,l/2}},
\label{eq:Fr} 
\end{align}
with $g$ being the gravitational acceleration. Eq.~\eqref{eq:KBC} is known as the kinematic boundary condition and Eq.~\eqref{eq:DBC} is known as the dynamic boundary condition of the free surface. The derivations of Eqs.~\eqref{eq:POT}-\eqref{eq:DBC} are given in Appendix A.

The velocity potential, $\Phi$, also satisfies the no-penetration boundary condition on the surface of the piezoelectric plate \citep{alben_flapping-flag_2008,alben_optimal_2008},
\begin{align}    
\pard{Y}{t}\,+\frac{\p Y}{\p x} &=\frac{\p \Phi}{\p y},\,\,\,\,\,\,\,\, {\text{at}\,\,\, y=0, }  \label{eq:BBC} 
\end{align}
and finally, the flow velocity should be finite at the trailing edge. To achieve this, a free vortex is shed from the trailing edge of the plate and the strength of the newly shed vortex is determined from the Kutta condition at the trailing edge of the piezoelectric plate \cite{saffman1995vortex}.

Assuming a harmonic incoming wave, we can express $\Phi$ as a summation of two harmonic potential functions,
\begin{align}    
\Phi\left(x,y,t\right) &=\Re\left[\left(\phi_0+\phi_1\right)e^{\ci \omega t}\right],  \label{eq:Phidecompose} 
\end{align}
{where $\Re$ implies the real part of the equation. Here, $\phi_0$ and $\phi_1$ are functions of spatial coordinates only where   $\phi_0$ is the known harmonic potential function of the incident wave with the non-dimensional angular frequency of $\omega$, and $\phi_1$ is velocity potential due to disturbed flow as a result of the inclusion of the piezoelectric plate in the flow field. We can further divide $\phi_1$ into three components, $\phi_1=\phi_1^W\,+\phi_1^I\,+\phi_1^F$, where $\phi_1^W$ is the contribution to the potential from the plate and its wake, $\phi_1^I$ is the contribution from s image system of $\phi_1^W$ outside the free surface, and $\phi_1^F$ is the contribution from the free surface when there is no incoming wave. In particular, in order to satisfy the free surface conditions in our problem, the image system consists of a vortex sheet  identical to vortex sheets in the plate and its wake 
is assumed at $h$ above the free surface.  The
combined effect of the real vortex sheet and its image ensures that $\phi_1^W\,+\phi_1^I\,=0$ on the mean free surface and therefore, $\phi_1^F$ is the primary function that describes the perturbed free surface.

In this study, we only consider incoming waves propagating in the positive $x$-direction. These waves are known as the head wave, while if the incident wave propagates opposite to the current direction, it is known as the following wave \citep{grue1985wave}.} The incoming head wave elevation $\eta_0$ is expressed as,
\begin{align} 
\eta_0\,=\, A_0 \sin\left(\omega t -k_0 x\right),
\label{eq:IncomWave} 
\end{align}
where $A_0$ is the non-dimensional amplitude of the incoming wave, and $k_0$ is the wavenumber of the admissible head wave \citep{haskind1954wave},
\begin{align} 
k_0=\frac{1}{2 Fr^2}\left[1+2 \omega Fr^2\,- \sqrt{1+4\omega Fr^2}\right],
\label{eq:k_0} 
\end{align}
and $\phi_0$ can be expressed as,
\begin{align} 
\phi_0(x,y)\,=\,\frac{A_0}{ \sqrt{k_0} Fr} e^{k_0 (y-h)}e^{-\ci k_0 x}. \label{eq:phi_0}
\end{align}

Since the incident wave and consequently the motion of the body are harmonic, the vortex strength along the body and in the wake can be expressed as  $\Lambda\left(x,t\right)=\Re\left[\gamma(x) e^{\ci \omega t}\right]$, allowing the non-dimensional $\phi_1^W$ and $\phi_1^I$ to be related to $\gamma(x)$ through \citep{crimi1964forces},
 \begin{align} 
\phi_1^W(x,y)&=-\frac{1}{2\pi}\int_{-1}^1 \gamma(x') \tan^{-1}\left(\frac{y}{x-x'}\right)\diff x'-\frac{\ci 
\omega \bar{\Gamma}}{2\pi}\int_{1}^\infty e^{-\ci \omega(x'-1)}\tan^{-1}\left(\frac{y}{x-x'}\right)\diff x', \label{eq:phi_1pW} \\
\phi_1^I(x,y)&=-\frac{1}{2\pi}\int_{-1}^1 \gamma(x')  \tan^{-1}\left(\frac{y-2h}{x-x'}\right)\diff x'-\frac{\ci 
\omega \bar{\Gamma}}{2\pi}\int_{1}^\infty e^{-\ci \omega(x'-1)}\tan^{-1}\left(\frac{y-2h}{x-x'}\right)\diff x', \label{eq:phi_1pI}
\end{align}
where $\bar{\Gamma}$ is the harmonic amplitude of total circulation of the plate (\emph{i.e.} $\Gamma=\Re\left[\bar{\Gamma}e^{\ci\omega t}\right]$) and is related to $\gamma$,
 \begin{align}
 \bar{\Gamma}\,=\int_{-1}^{1}\gamma(x')\diff x'. \label{eq:Gamma}
\end{align}

Finally $\phi_1^F(x,y)$ can be expressed as, 
\begin{align} 
\phi_1^F(x,y)&=\frac{\ci}{2\pi Fr^2}\int_{-1}^1 \gamma(x') G(x';x,y)\diff x'+\frac{
\omega \bar{\Gamma}}{2\pi Fr^2}\int_{1}^\infty e^{-\ci \omega(x'-1)}G(x';x,y)\diff x', \label{eq:phi_1pF}
\end{align}
where $G(x';x,y)$ is the fundamental solution for an oscillating point vortex beneath the free surface \citep{tan1955source,tan1957waves} and is given by,
\begin{align}
G(x';x,y)\,&=\int_0^\infty e^{\sigma(y-2h)}
\left[\frac{e^{\ci\sigma(x-x')}}{(\sigma-\sigma_1)(\sigma-\sigma_2)}\,-
        \frac{e^{-\ci\sigma(x-x')}}{(\sigma-\sigma_3)(\sigma-\sigma_4)}\right]\diff \sigma \nonumber\\
               &+D_1\, e^{\sigma_1\,\left[y-2h+\ci(x-x')\right]}
                 +\,D_2\, e^{\sigma_2\,\left[y-2h+\ci(x-x')\right]}
                 \nonumber\\
                &+D_3\, e^{\sigma_3\,\left[y-2h-\ci(x-x')\right]}
                 +\,D_4\, e^{\sigma_4\,\left[y-2h-\ci(x-x')\right]}, \label{eq:G}
\end{align}
where 
\begin{align}
\sigma_{1,2}\,&=\frac{1}{2 Fr^2}\left[1-2\omega Fr^2\,\pm \sqrt{1-4\omega Fr^2}\right],\nonumber\\
\sigma_{3,4}\,&=\frac{1}{2 Fr^2}\left[1+2\omega Fr^2\,\pm \sqrt{1+4\omega Fr^2}\right].\label{eq:sigma}
\end{align}

Here, $\sigma_{1}$ to $\sigma_4$ are the wavenumbers of possible wave trains created as a result of the insertion of the plate. $D_{1}$ to $D_4$ are complimentary coefficients which are calculated such that $G(x';x,y)$ provides the correct radiation free surface condition at $x \rightarrow \pm \infty$. Further information can be found in \cite{reece_motion_nodate}.

 The relations of $\phi_1^W$, $\phi_1^I$ and $\phi_1^F$ to the vortex strength along the body and in the wake, $\gamma$ (Eqs.~ \ref{eq:phi_1pW}, \ref{eq:phi_1pI} and \ref{eq:phi_1pF}) can be used to express the no-penetration boundary condition (Eq. \ref{eq:BBC}) along the plate ($-1\le x\le 1$) based on $\gamma$ as,
\begin{align}
-\ci \omega {\xi(x)} &- \pard{\xi(x)}{x}+\left.\pard{\phi_0(x,y)}{y}\right|_{y=0}=\frac{1}{2\pi} \Pint_{-1}^1 \frac{\gamma(x')}{x-x'}\diff x' \nonumber \\
&+\frac{1}{2\pi} \int_{-1}^1 \gamma(x')\left[\frac{x-x'}{(x-x')^2+(2h)^2}-\frac{\ci}{Fr^2}G_y(x') \right]\diff x' \nonumber \\
&-\frac{\ci\omega\bar{\Gamma}}{2\pi} \int_{1}^\infty e^{-\ci \omega(x'-1)}\left[\frac{1}{x-x'}+\frac{x-x'}{(x-x')^2+(2h)^2}-\frac{\ci}{Fr^2}G_y(x') \right]\diff x', \label{eq:gamma}
\end{align}
where $Y\left(x,t\right)=\Re\left[\xi(x) e^{\ci \omega t}\right]$ and $G_y(x') =\left.\p G(x';x,y)/\p y \right|_{y=0}$. $\Pint$ is used for the Cauchy principle value integral. Similarly, other equations can also be written based on the normal modes of time-periodic variables.  Using the notation of $[P](x,t)=\Re\left[[p](x)e^{\ci\omega t}\right]$ and $V(x,t)=\Re\left[v(x)e^{\ci\omega t}\right]$, Eqs.~\eqref{eq:EOM}, \eqref{eq:EOE} and \eqref{eq:EOP} are combined into,
\begin{align}
\left(1+\frac{\alpha^2 \beta \omega^2}{\beta\omega^2-\ci\omega-\beta\tau^2}\right)\,\frac{\p^4 \xi}{\p x^4}\, &=\, -{{U^*}^2}\,[p], \label{eq:EOMw} \\
\ci\omega{\gamma}\,+\, \pard{\gamma}{x}\, &=\,-\pard{[p]}{x}, 
\label{eq:EOPw}
\end{align}
with the boundary conditions,
\begin{align}
\xi=\,  \pard{\xi}{x}\, =\, 0\,\,\,\, &\text{ at } x=-1,\label{eq:BCfixedw}\\
\pardo{\xi}{x}=\,\frac{\p^3 \xi}{\p x^3}=\, 0\,\,\,\, &\text{ at } x=\,\,\,\, 1, \label{eq:BCfreew}\\
\text{ Finite velocity (Kutta condition)} &\text{ at } x=\,\,\,\, 1. \label{eq:BCprew}
\end{align}

The system of Eqs. \eqref{eq:gamma}-\eqref{eq:EOPw} along with boundary conditions Eqs.~\eqref{eq:BCfixedw}-\eqref{eq:BCprew} are solved for $\xi$, $[p]$ and $\gamma$. 
The electrical power output from the piezoelectric plate is, 
\begin{align}
\overline{\dot{W}_e}=\frac{1}{\beta}\frac{\omega}{2\pi} \int_0^{\frac{2\pi}{\omega}} \int_{-1}^1 V^2(x,t) \diff x \diff t = \frac{1}{2\beta} \int_{-1}^1 \left|v(x)\right|^2 \diff x.
\label{eq:We}
\end{align}
The non-dimensional power input from the incoming wave is calculated from {Eq. \eqref{eq:Ww}} for the wave train progressing in the direction of $u$, {commonly referred to as the "head wave."}
\begin{align}
\overline{\dot{W}_w}= E_w \left|c_g+1\right|,
\label{eq:Ww}
\end{align}

In the above relation, $E_w = \frac{1}{2}A_0^2/Fr^2$ is the incoming wave energy density, and $c_g =  1/\left(2\sqrt{k_0}Fr\right)$ is the group velocity of the head wave. In addition, the mean thrust force acting on the piezoelectric plate, $\overline{T}$, can be computed from
\begin{align}
\overline{T}\,=\, \overline{T}_p + \overline{S},
\label{eq:D}
\end{align}
where $\overline{T}_p$  is the thrust force due to the pressure difference along the plate given by,
\begin{align}
   \overline{T}_p=\frac{\omega}{2\pi} \int_0^{\frac{2\pi}{\omega}} \int_{-1}^1 [P](x,t)\, \frac{\p Y}{\p x}(x,t) \diff x \diff t = \frac{1}{2} \int_{-1}^1 [p](x)\, \xi^*(x)\ \diff x.
   \label{eq:D_p}
\end{align}

Here, $\xi^*$ is the complex conjugate of $\xi$. $\overline{S}$ is a non-dimensional suction force at the sharp leading edge of the plate, pointing toward minus $x$ direction. The suction force is calculated by applying the Blasius formula to a small circle $R_{\delta_0}$ of radius $\delta_0$ surrounding the leading edge. In particular, we can write $S=  \frac{1}{2}\oint_{R_{\delta_0}} {\Psi'}^2 (z) \diff z$, where $z=x+\ci y$ is the complex number represents the position in $x-y$ plane, and $\Psi'(z)=u-\ci v$ is the complex velocity \citep{saffman1995vortex}. 
   
The mean energy equation, averaged in time, in its non-dimensional form can be expressed as
\begin{align}   
\overline{\dot{W}_w}=\overline{\dot{W}_e}+\overline{T}+\overline{E},
\label{eq:energy}
\end{align}
where $\overline{E}$ is the summation of mean wasted energy from the  wake of the plate and scattering surface waves from the inclusion of the plate below the free surface. To quantify the overall efficiency of the piezoelectric plate, we define
the energy harvesting efficiency $\eta_P$,
\begin{align}   
\eta_P\,&=\,\,\,\,\frac{\overline{\dot{W}_e}}{\overline{\dot{W}_w}}.
\label{eq:eta_P}
\end{align}

\subsection{Numerical method}
 For the numerical simulation,  $m+1$ Chebyshev-Lobatto collocation points are chosen to discretize  $\frac{\p^2\xi}{\p x^2}$ along the plate,
\begin{align}
x_j &=-\cos\,\left(\frac{j\pi}{m}\right) ,\qquad \textrm{with       }\,\,  j=0,\cdots,m.\label{eq:chebyshev} 
\end{align}

Knowing $[p]$ at the $j$th collocation point, $\frac{\p^2\xi}{\p x^2}$ can be easily computed from Eq. \eqref{eq:EOMw} along with the boundary conditions at the free end of the plate (Eq. \ref{eq:BCfreew}). Then, by integrating $\frac{\p^2\xi}{\p x^2}$ twice and incorporating the
boundary conditions at the fixed end (Eqs. \ref{eq:BCfreew} and \ref{eq:BCfixedw}), the plate deflection $\xi(x_j)$ can be computed. To solve Eq. \eqref{eq:gamma},  we employ a Glauert series solution for  $\gamma$,
\begin{align} 
\gamma(\theta)=2 \left[ a_0 \cot\frac{\theta}{2}+\sum_{j=1}^{\infty} a_j \sin\left(j\theta\,\right)\right],
\label{eq:Glauret}
\end{align}   
with $\theta=-\text{arccos} (x)$ and rewrite Eq. \eqref{eq:Glauret} as,
\begin{align} 
\frac{1}{2\pi} \Pint_{-1}^1 \frac{\gamma(x')}{x-x'}\diff x' = \left[a_0 - \sum_{j=1}^{\infty} a_j \cos \left(j\theta\,\right) \right].
\label{eq:gammacos}
\end{align} 
  
We employ Fourier cosine series representations of variables and group the coefficients of $\cos\left(j\,\theta\right)$ terms to obtain an infinite set of equations to solve for the coefficients. In particular, the governing equation for $\gamma$ (Eq. \ref{eq:gamma}) can be written as the set of algebraic equations for $a_i$,
\begin{align} 
a_i-\sum_{j=0}^{\infty} Q_{ij} a_j = r_i,
\label{eq:a_0}
\end{align} 
where $Q_{ij}$ coefficients account for the second and third terms in the right side of Eq. \eqref{eq:gamma} as highlighted in \cite{crimi1964forces}. 
In Eq. \eqref{eq:a_0}, $r_i$ is  obtained from the cosine Fourier expansion coefficients of the left-hand side of Eq. \eqref{eq:gamma} and can be computed from,
\begin{align}
r_i=\left[C^{-1}\right]_{ik}\,\, \text{LHS}\left(x_k\right),
\label{eq:r_i}
\end{align}
where $\text{LHS}\left(x_k\right)$ is the summation of all terms in the left-hand side of Eq. \eqref{eq:gamma} at the collocation point $x_k$ along the plate.  The transformation matrix, $C_{ij}$x, is defined as,
\begin{align}
C_{ij}=\left\{ 
   \begin{array}{ll}
         1  & \mbox{if $j = 0$};\\
       -\cos \left(\frac{ij\pi}{m}\right) & \mbox{if $j \geq 1$}. 
   \end{array} \right. 
   \label{eq:C_ij}
\end{align}

Similarly, upon inserting Eq. \eqref{eq:Glauret} in Eq.  \eqref{eq:EOPw}, the pressure distribution on the plate can be expressed as,
\begin{align}
[p]=-2 \left[ a_0 \cot\frac{\theta}{2}+\sum_{j=1}^{\infty} b_j \sin\left(j\,\theta\,\right) \right]
\label{eq:pseries}
\end{align}
where 
\begin{align}
 b_j=\left\{ 
\begin{array}{ll}
         a_1+\ci\omega \left[3a_0+a_1+\frac{a_2}{2}\right]  & \mbox{if $j = 1$};\\
        a_j+\frac{\ci\omega}{2j}\left[a_{j+1}-a_{j-1}+2\left(2a_0+a_1\right)\left(-1\right)^{j+1}\right] & \mbox{if $j \geq 2$}. \end{array} \right. 
 \label{eq:bj}       
\end{align}

It appears from Eq. \eqref{eq:pseries} that $\left[p\right]$ is singular at the leading edge $x=-1$, and therefore from Eq. \eqref{eq:EOMw}, $\frac{\p^4\xi}{\p x^4}$ is singular the leading edge.
This
does not pose a problem as  Eq. \eqref{eq:EOMw} is only evaluated for the internal nodes. In particular, we solve the second order Eq. \eqref{eq:EOMw} for $\xi_{xx}$ and replace the equations on the boundary nodes of $x_0$ and $x_m$ with the two boundary conditions at the free end of the plate, $\xi_{xx}(x_m)=\xi_{xxx}(x_m)=0$, to solve the following system of $m+1$ equation,
\begin{align}
         \left(1+\frac{\alpha^2 \beta \omega^2}{\beta\omega^2-\ci\omega-\beta\tau^2}\right)\, D_{xx}\, \xi_{xx}\, &=\, -{U^*}^2[p]  \mbox{\,\,\,\,\,\, for \,\,\,\, $j = 1,\ldots,m-1$},\\
       \xi_{xx}\left(x_m\right)\, &=\,\, 0,   \\
       D_x\, \xi_{xx}\left(x_m\right)\, &=\,\, 0, 
 \label{eq:EOMw2}
\end{align}
where $D_x$ and $D_{xx}$ are the first and second order Chebyshev differentiation matrices. 
Following the convergence study, 21 Chebyshev-Lobatto nodes along the plate length are found to be sufficient to discretize the governing equations. Similarly, the series expansions in Eqs. \eqref{eq:Glauret} and \eqref{eq:pseries} are also truncated to contain only the first 21 terms. 

\section{Validation}
\begin{figure}
			\centering{
		\begin{subfigure}{0.58\textwidth}
		\centering  \includegraphics[width=\linewidth]{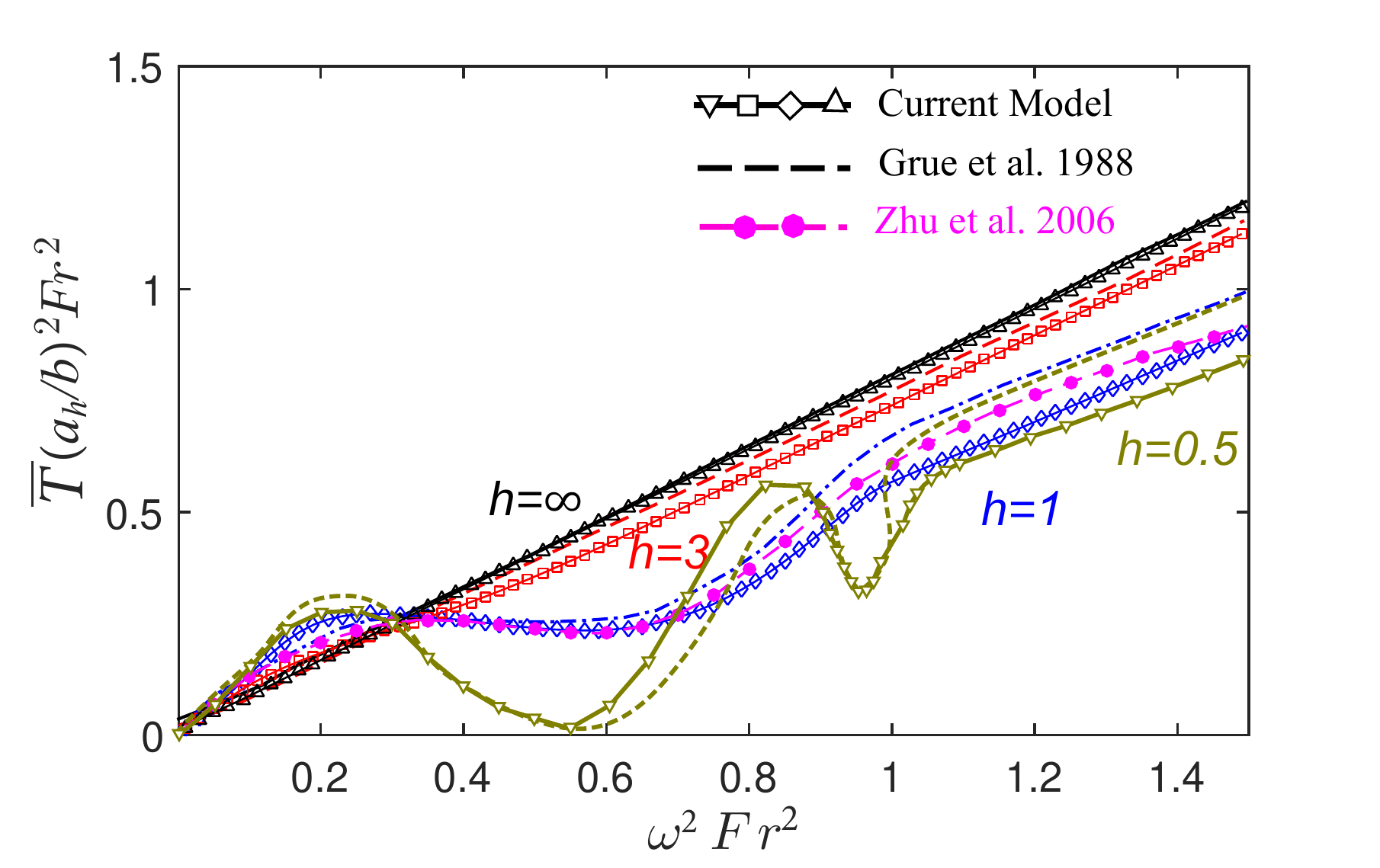}
		\end{subfigure} \hspace{-0.7cm}
	    \begin{subfigure}{0.40\linewidth}
	    \centering 
        \includegraphics[width=\linewidth]{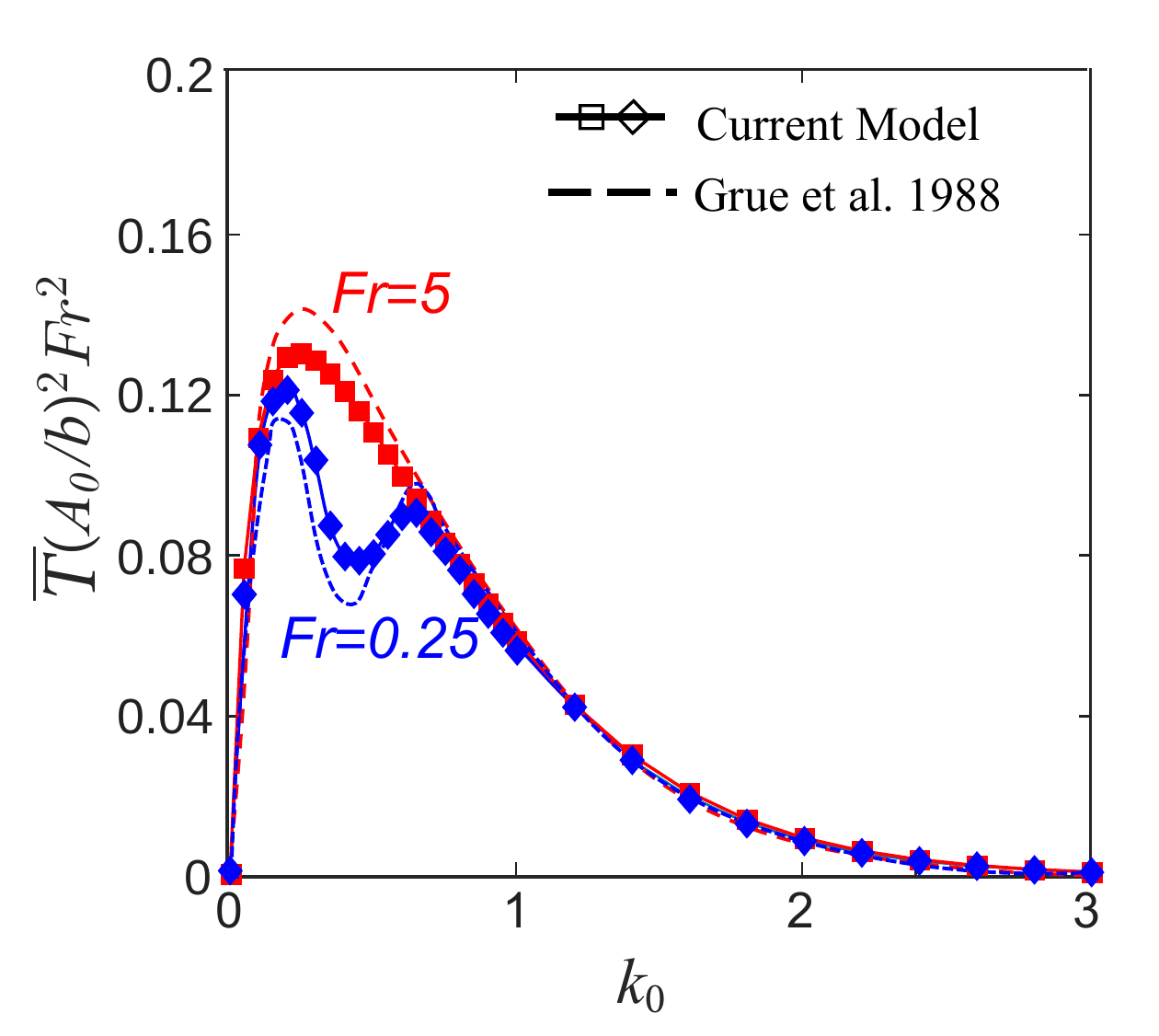}
		\end{subfigure}
		\put(-480,180){$(a)$}
		\put(-200,180){$(b)$}
		}
		\caption{a) Dependence of thrust force on the heaving frequency of  a horizontal plate below the free surface. $Fr=0.25$ and $h=0.5$, 1, 3, $\infty$.  Dashed lines are for \cite{grue_propulsion_1988}; magenta circular symbols are the numerical prediction for \cite{zhu2006dynamics} obtained for $a_h/b=0.05$ with $a_h$ being the heaving amplitude. The Other symbols are for the current model. b) The comparison of the thrust force for different incident wavenumbers between the current model (symbols) and \cite{grue_propulsion_1988} (dashed lines) for a fixed plate with $h=1$ and $Fr=0.25$ and 5. Here $A_0$ and $k_0$ are the non-dimensional amplitude and wavenumber  of the incident wave.}
		\label{figValidation}
\end{figure}

{To validate the present model, we compare the thrust force of the plate with \cite{grue_propulsion_1988} for the scenarios in which the plate undergoes periodic heaving motion close to the calm free surface without any incident wave in Fig. \ref{figValidation}a. Four submergence depths of $h=0.5, 1,3,\infty$ are tested while the Froude number is fixed at $Fr=0.25$ for all cases. The non-dimensional amplitude of heaving motion is denoted with $a_h/b$. In addition, the numerical results of the same problem for a three-dimensional foil beneath the free surface by \cite{zhu2006dynamics} are also included. In \cite{zhu2006dynamics}, the heaving motion of a thin NACA0005 foil with a large span-to-chord length ratio of 10 and a small heaving amplitude of $a_h/b=0.05$ is simulated for the submergence depth of $h=1$. {The large span-to-chord ratio of this numerical results creates a practically two-dimensional flow field similar to the conditions examined here}. The present result compares well with previous analytical and numerical results. The maximum difference for $h<1$ cases is less than $3\%$, while the maximum difference of $15\%$ is seen between our prediction and reported values in \cite{grue_propulsion_1988} for high frequency and small $h=0.5$ cases.     

In addition, we cross-compare our predictions with the reported values of a fixed plate exposed to the incident waves, as shown in Fig. \ref{figValidation}b. Here, the plate is kept fixed at its initial position and the thrust forces for different incoming wavenumbers are compared to reported values in \cite{grue_propulsion_1988}. The submergence depth is $h=1$ and two $Fr=0.25,5$ are tested. The prediction from the current method closely follows the previously report values with a maximum error of less than $10\%$. }  
\section{Results}

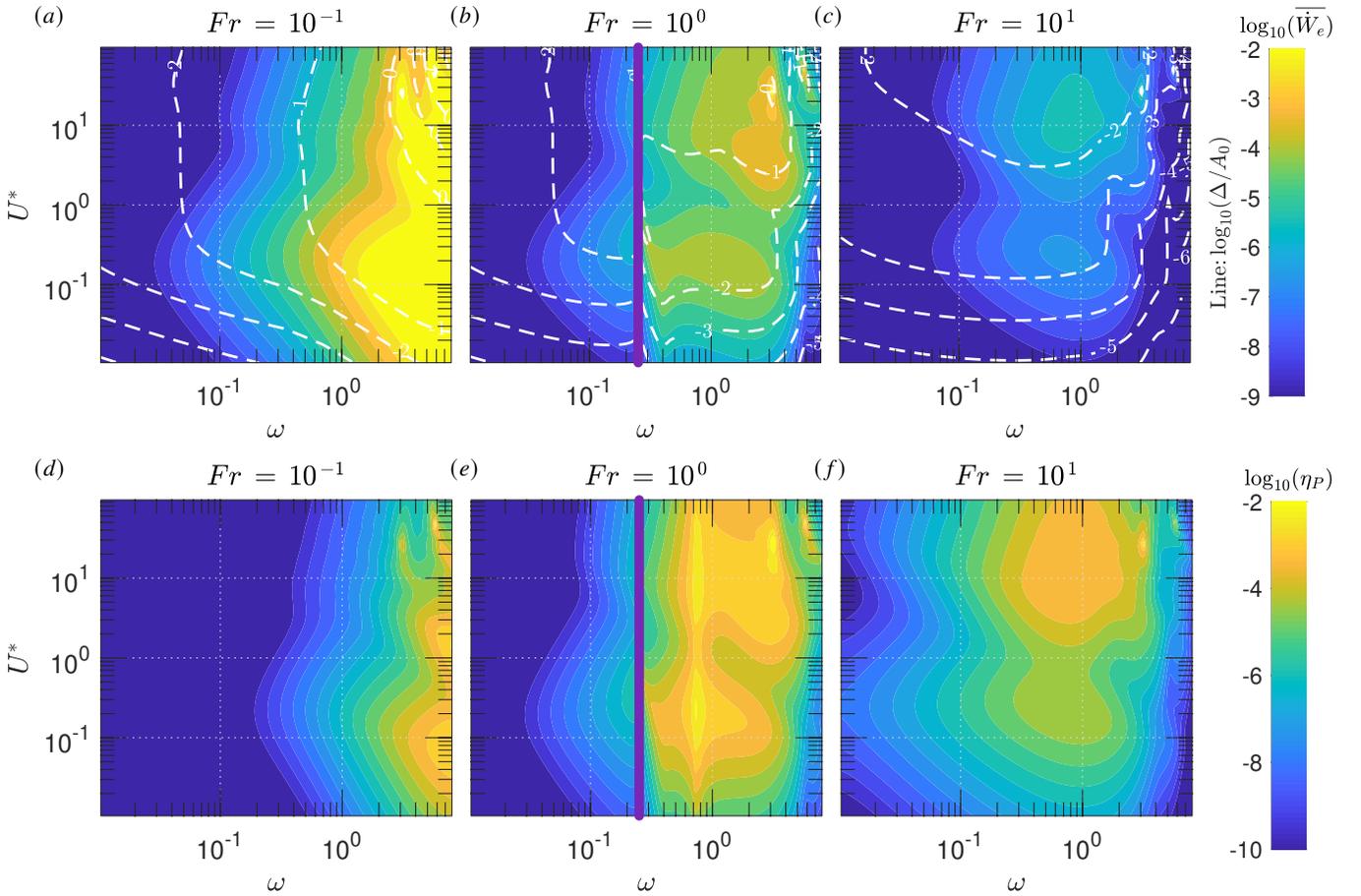
\begin{figure}
\centering{
\includegraphics[width=\textwidth]{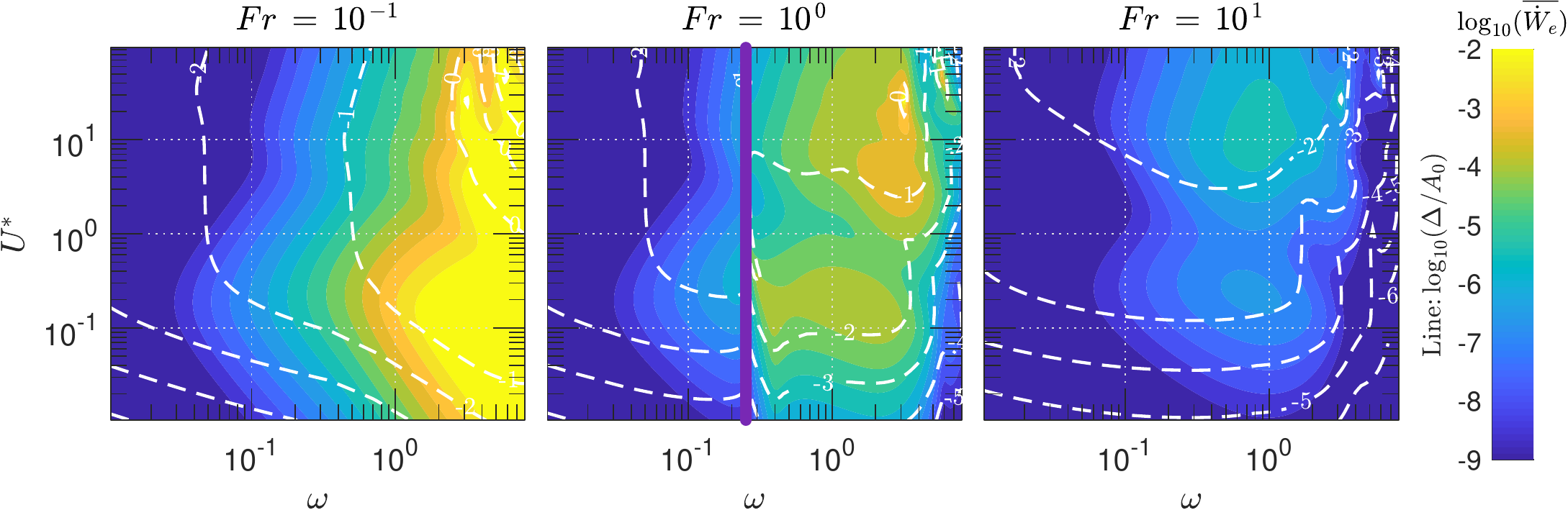}
		\put(-500,160){$(a)$}
        \put(-340,160){$(b)$}
        \put(-200,160){$(c)$}\\ \vspace{0.25cm}
\includegraphics[width=\textwidth]{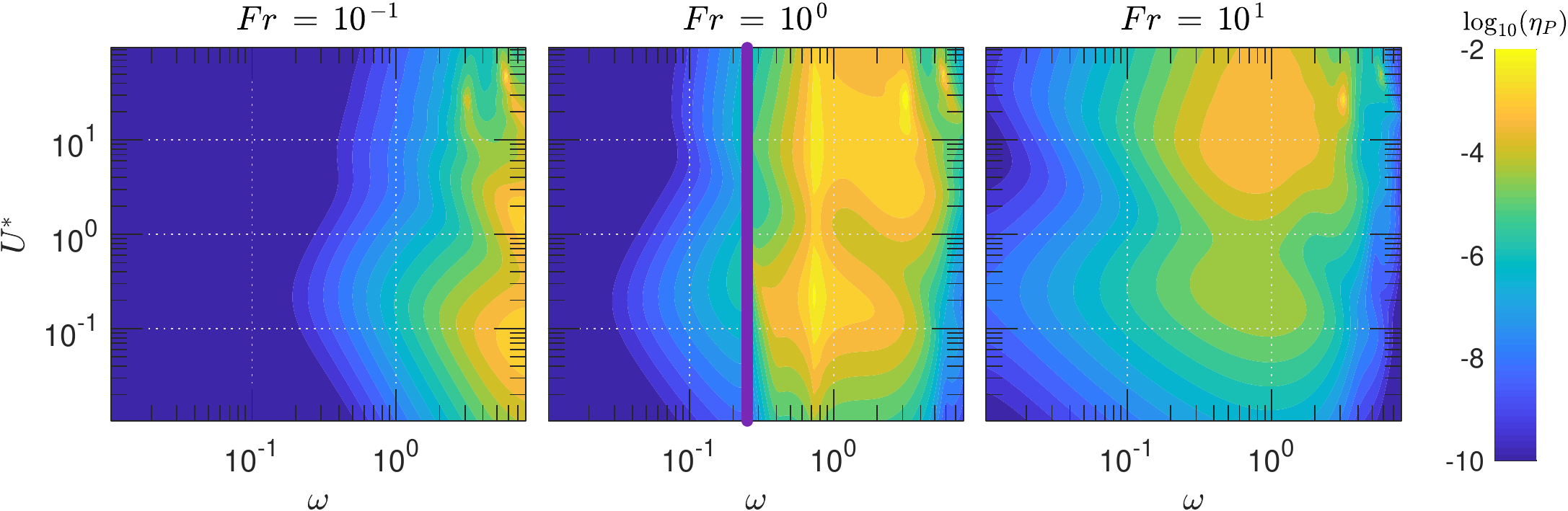} 		
\put(-500,160){$(d)$}
        \put(-340,160){$(e)$}
        \put(-200,160){$(f)$}
        }
\caption{(a-c) The average electrical power output and lateral deflection amplitude (lines)  and (b-d) energy efficiency  versus $\omega$ and $U^*$ for three representative $Fr=0.1$ (a,d), 1 (b,e) and 10 (c,f). Other parameters are fixed at $\alpha=0.5$, $\beta=1$, $\tau=0$, and $h=1.0$. The regions close to $\omega Fr^2=\frac{1}{4}$ is masked with a purple line since the proposed linear solution is not valid due to abrupt changes of surface wave propagation modes \citep{palm1999wave}. 
\label{Fig1_SLIDE24}  }
\end{figure}

In the following, we investigate how the characteristic parameters influence the vibration behavior and energy harvesting capabilities of the piezoelectric plate energy harvester.
\subsection{Effects of Reduced Velocity}
The bending rigidity of a piezoelectric plate is characterized using the non-dimensional reduced velocity, $U^*$, presented in Eq. \eqref{eq:nondim}. In Fig. \ref{Fig1_SLIDE24}a-c, the electrical power output $\overline{\dot{W}_e}$ and the ratio of maximum lateral deflection of the plate to the incident wave amplitude $\Delta/A_0$ are plotted for a wide range of $10^{-2}<U^*<10^2$ and $10^{-2}<\omega<10$ for three representative Froude numbers of $Fr=0.1$ (subcritical flow), $Fr=1$ (critical flow), and $Fr=10$ (supercritical flow). {Here, $\Delta$ is defined as the peak-to-peak lateral deflection of the plate defined as $\Delta=y_{\max}-y_{\min}$, where $=y_{\max}$ and $y_{\min}$ are the maximum and minimum lateral positions of the plate.} The immediate regions of $\omega Fr^2=\frac{1}{4}$ are removed from the plots with the purple line. At this condition, the radiative waves originating from the plate switch from  $\sigma_1$ to $\sigma_4$ waves to only include $\sigma_3$ and $\sigma_4$ waves and hence, the proposed linear solution stops being valid \citep{palm1999wave}. All other parameters are fixed at their representative values of {$h=1.0$}, $\alpha=0.5$, $\beta=1$, and $\tau=0$.   

For the subcritical flow, the output energy is mainly impacted by the wave frequency and only marginally changes with the flexibility of the plate represented with $U^*$. Nonetheless, for higher frequencies, two distinct maxima are observed for all $Fr$ cases, one at $U^*>1$ resembling the flow-induced flapping mode of the plate and the other one at $U^*<1$ related to the cantilever resonance mode of the plate, hereafter is referred to as the fluttering mode. Looking at subcritical flow with $Fr=0.1$ in Fig. \ref{Fig1_SLIDE24}a, we can see that while $\overline{\dot{W}_e}$ decays quickly as $\omega$ decreases below 1, the maximum deflection of the plate is almost independent of  $U^*$  in this range if the plate vibrates with the flapping mode. On the other hand, contour lines with similar $\Delta/A_0$ have a linear trend in $\log_{10}(\omega)$-$\log_{10}(U^*)$ plane if the plate oscillates in its fluttering mode. The exception is an isolated small region at higher $U^*$ and $\omega \approx 2$  where there is a sudden increase in the flapping amplitude and the electrical energy output. The same observation can be made across different $Fr$ numbers wherein the maximum harvested energy is localized at a narrow range of frequencies and $U^*$. In this condition, the convective timescale of the problem and the encountered wave frequency take a particular ratio that promotes better energy transfer from the flow to the plate. 

The energy harvesting efficiencies, $\eta_P$ , for $Fr=0.1, 1, 10$  are shown in Fig.  \ref{Fig1_SLIDE24}d-f.  The energy harvesting efficiency increases with $\omega$ when $Fr$ is small and the flow is critical or subcritical. When $\omega Fr^2<\frac{1}{4}$, the energy harvesting efficiency shows larger sensitivity to $\omega$ compared to $U^*$. On the other hand, when $\omega Fr^2>\frac{1}{4}$ (the right side of the purple line of $Fr=1$ cases and  the entire region of $Fr=10$ cases in Fig. \ref{Fig1_SLIDE24}b), $\eta_P$ is affected by both  $U^*$ and $\omega$. In this case, {only one new surface wave $\sigma_2$ will be generated in addition to the incident head wave, $\sigma_4$, and both of these waves propagate downstream.} The incoming wave partially transmits to the electrical energy while the rest of the energy propagates in the downstream direction as new types of waves.

\begin{figure}
\includegraphics[width=1\textwidth]{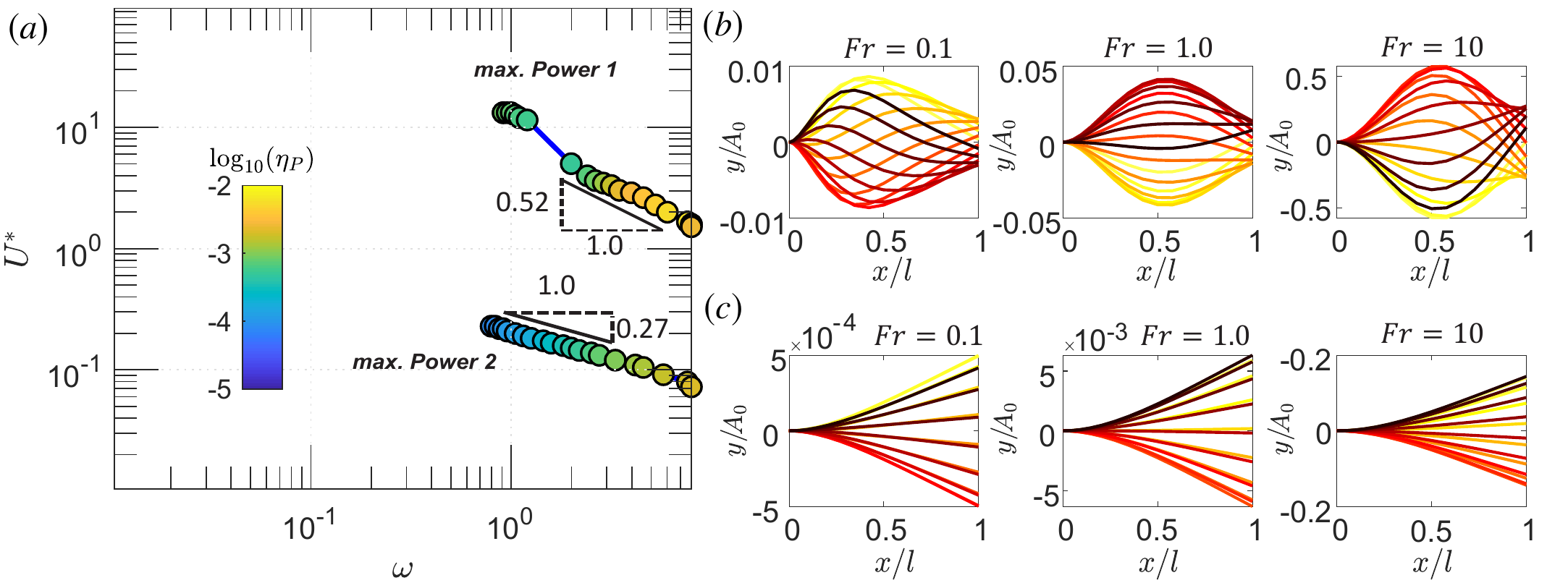}
\caption{a) The location of the most energy efficient condition in $\omega - U^*$ parametric space with the changes of $Fr$.   b) the snapshot of the plate deflection over one cycle along the first branch (flapping case) with time instance is marked from light to dark color for a unit incoming wave amplitude. c) the mode of flutter along the second branch.
\label{Fig2_SLIDE25MAX}  }
\end{figure}

 Figure \ref{Fig2_SLIDE25MAX}a shows the locations of the two highest values of $\eta_P$ in the $\omega$-$U^*$ plane for different $Fr$ numbers. The first branch is located at high $U^*>1$ values, and the second branch is observed at $U^*<1$. The highest achievable $\eta_P$ is always observed for higher $Fr$ cases while its magnitude along both branches remains close. The position of the optimal case in the first branch follows a power law with an exponent of $\approx -0.52$ and shifts to lower $U^*$ and higher $\omega$ regimes with the decrease of $Fr$. The changes along the second branch are less dependent on the $U^*$ with the power-law exponent of $-0.27$. The plate undergoes different modes of vibration along each of these branches. The plate exhibits a classical travel wave deflection pattern reminiscent of the flag flapping problem along the first branch. Yet, the plate vibrates at much smaller $U^*$ compared to $U^*$ associated with the fluttering instability of light flags \citep{alben_flapping-flag_2008}. In this case, with an increase of $Fr$, the place of maximum lateral deflection is relocated along the length while the vibration amplitude reduces (Fig. \ref{Fig2_SLIDE25MAX}b). On the other hand, the optimal cases along the other branch consistently vibrate in their fluttering modes, as shown in  Fig. \ref{Fig2_SLIDE25MAX}c. In this case, the interaction between the first natural frequency of immersed plate and the frequency of the propagating surface wave increases the oscillation amplitude and, consequently, results in a higher level of harvested energy. 
 
The results suggest that one can practically change the plate length to attain maximum energy harvesting efficiency in two distinct response modes, flapping and cantilever fluttering modes. In particular, from the definitions of non-dimensional parameters, we can show that $U^* \propto b^{3/2}$, $\omega \propto b^{-1}$ and $Fr \propto b^{-1/2}$ and therefore depending on the flow speed and structural parameters, the plate length can be controlled to place the system on the first optimal branch. Otherwise, the system can also be switched to the cantilever model to maintain maximum energy harvesting efficiency. Also, when the incoming wave spectrum is broadband,  the cantilever fluttering mode will continue delivering maximum attainable energy. On the other hand, to benefit from the flapping mode, it is necessary to perform a complex control of $U^*$, perhaps by adjusting the flow velocity or, equivalently, the advancement speed of the plate.

\subsection{Effects of Froude Number}

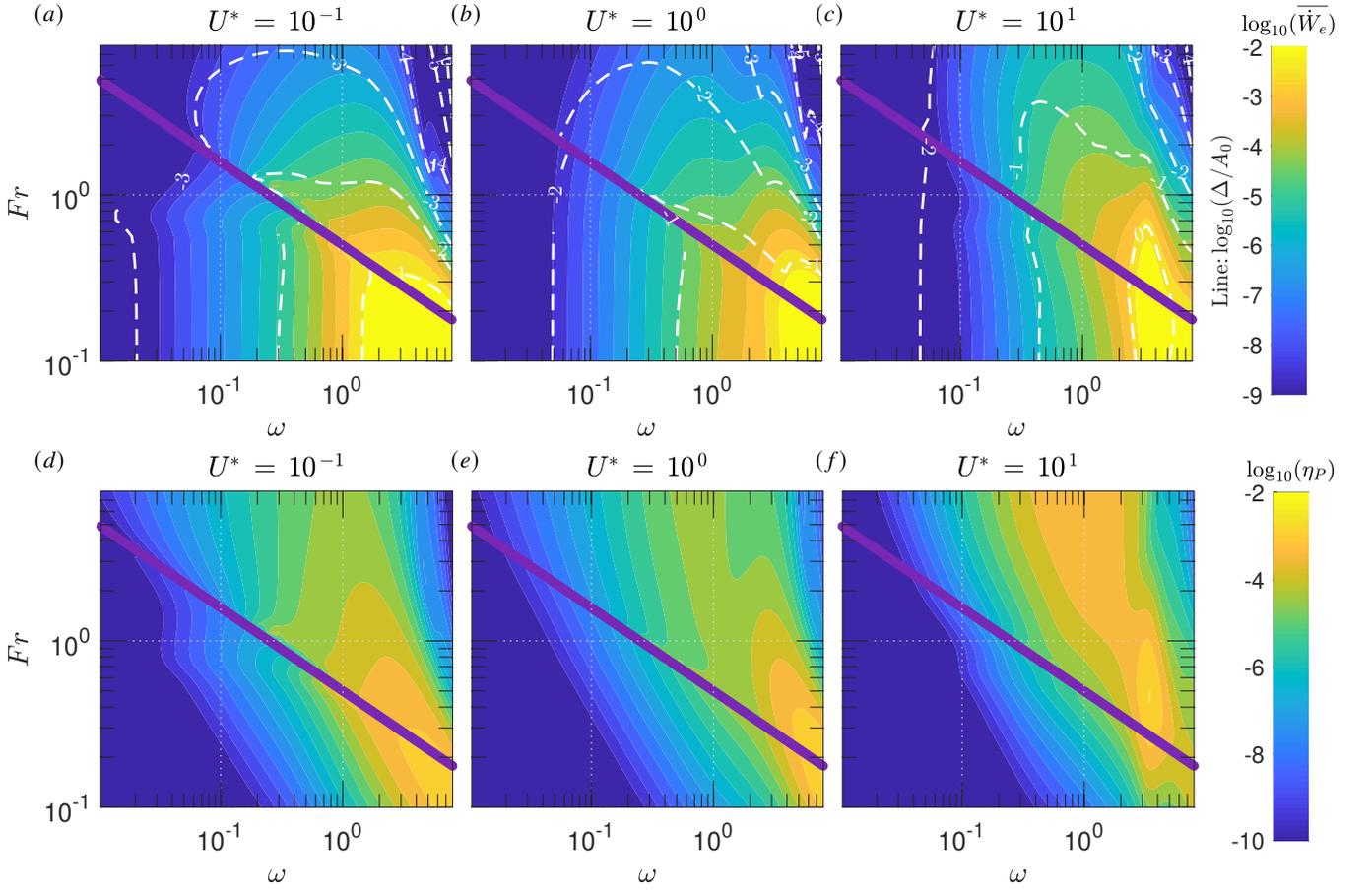
\begin{figure}
\centering{
\includegraphics[width=\textwidth]{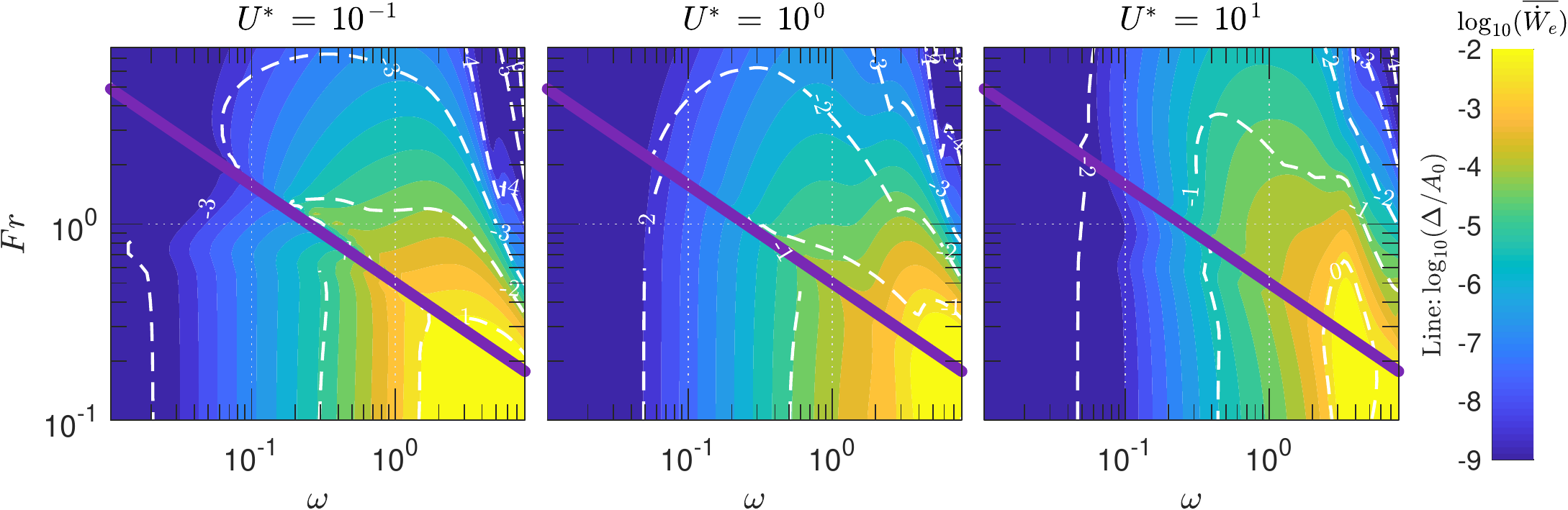}
		\put(-500,160){$(a)$}
        \put(-340,160){$(b)$}
        \put(-200,160){$(c)$}
        \\ \vspace{0.15cm}
\includegraphics[width=\textwidth]{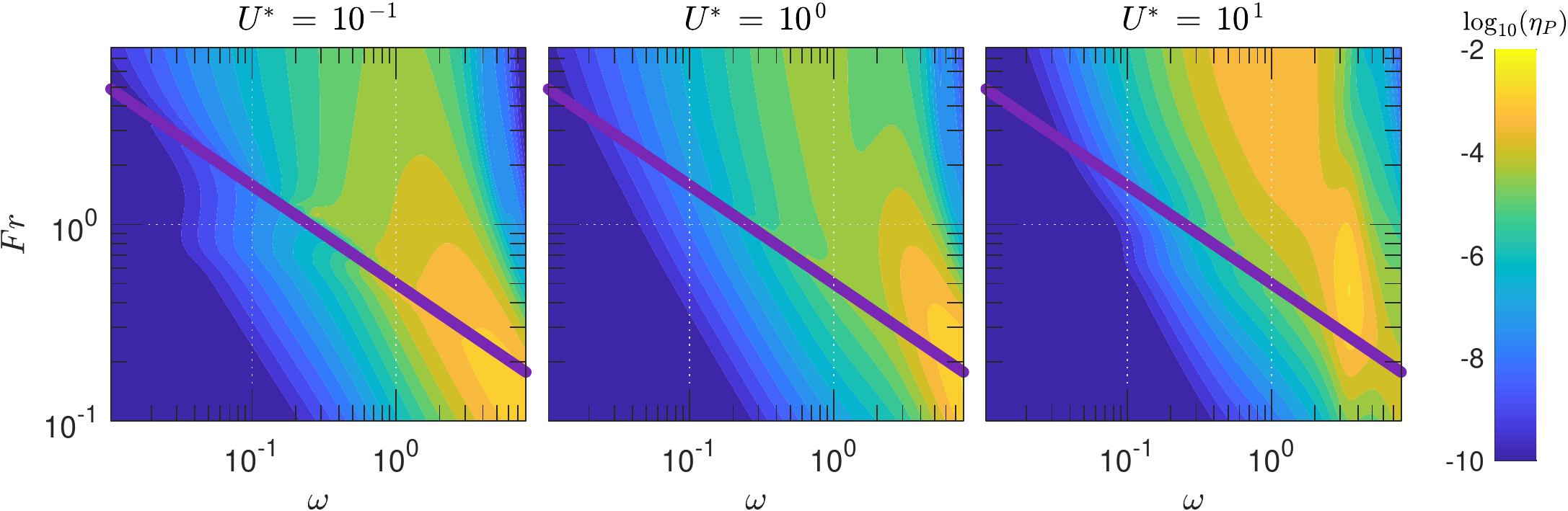}
		\put(-500,160){$(d)$}
        \put(-340,160){$(e)$}
        \put(-200,160){$(f)$}\\ \vspace{0.05cm}
}
\caption{ Dependency of 
$\overline{\dot{W}_e}$, $\Delta/A_0$ (a-c) and $\eta_P$,  (d-f) \emph{vs.} $\omega - Fr$ for three representative $U^*$ of $0.1, 1, 10$.
Other parameters are fixed at $\alpha=0.5$, $\beta=1$, $\tau=0$, and $h=1.0$. 
\label{Fig2_SLIDE26T}  }
\end{figure}

The interaction of the surface wave and the submerged plate is highly dependent on the Froude number, $Fr$. Figure \ref{Fig2_SLIDE26T} compares $\overline{\dot{W}_e}$ and $\Delta/A_0$ as a function of $Fr$ and $\omega$ for three representative reduced velocities of $0.1, 1$ and $10$ selected from the results discussed in the previous section. All other parameters are fixed at their representative values of $\alpha=0.5$, $\beta=1$, $\tau=0$, and $h=1$. For the stiff plate (Fig.   \ref{Fig2_SLIDE26T}a), below the critical line of $\omega Fr^2=\frac{1}{4}$, the harvested energy is mainly a function of $\omega$, while for the cases above the critical line, the harvested energy is primarily a function of $Fr$. In this case, the plate is actuated in its cantilever fluttering mode. Vibration amplitude follows a similar trend to $\overline{\dot{W}_e}$. At $U^*=1$, the vibration mode starts switching from fluttering to flapping mode and the region with higher $\overline{\dot{W}_e}$ extends to lower $\omega$ and higher $Fr$ cases (Fig. \ref{Fig2_SLIDE26T}b). 

For the very flexible plate with $U^*=10$ shown in Fig.  \ref{Fig2_SLIDE26T}c, both  $\overline{\dot{W}_e}$ and $\Delta/A_0$ are predominantly functions of $\omega$, and $Fr$-dependency is only observed  over a narrow region in the parametric space. In this case, the energy-producing region shrinks along the $\omega$ axis but occurs over a wider range of $Fr$ numbers. The vibration mode comprises traveling flapping modes with one, two, or three nodes along the plate length. 

The variations of $\eta_P$ with $\omega$ and $Fr$ are shown in Fig. \ref{Fig2_SLIDE26T}d-f. For lower $U^*$ cases, different from $\overline{\dot{W}_e}$, the region with large energy harvesting efficiency $\eta_P$ extends to higher $Fr$ ranges where the plate converts a higher percentage of the incoming wave to electrical energy.

A good correlation is identified between the amplitude of vibration and $\eta_P$. It is observed that the highest efficiency in these cases is directly related to the amplitude of the fluttering mode and the highest efficiency could be found over a range of $\omega$ just before the dominant vibration mode switching from the flapping to fluttering mode. This trend changes at larger reduced velocities where the highest efficiency is concentrated at a narrow band of $\omega$. The region with high $\eta_P$ and  $\overline{\dot{W}_e}$  coincides at $\omega\approx 3$, with much lower dependency to $Fr$. It is found that for flexible plates, the wave can not actuate the first fluttering mode due to a large mismatch between the natural frequencies of the plate and the incoming wave and instead triggers and amplifies the flapping response of the plate. As a result, the plate  simultaneously harvests the energy from the wave and  current and reaches much higher $\eta_P$.

\subsection{Effects of Submergence Depth}
\begin{figure} 
		\centering{
		\includegraphics[width=\textwidth]{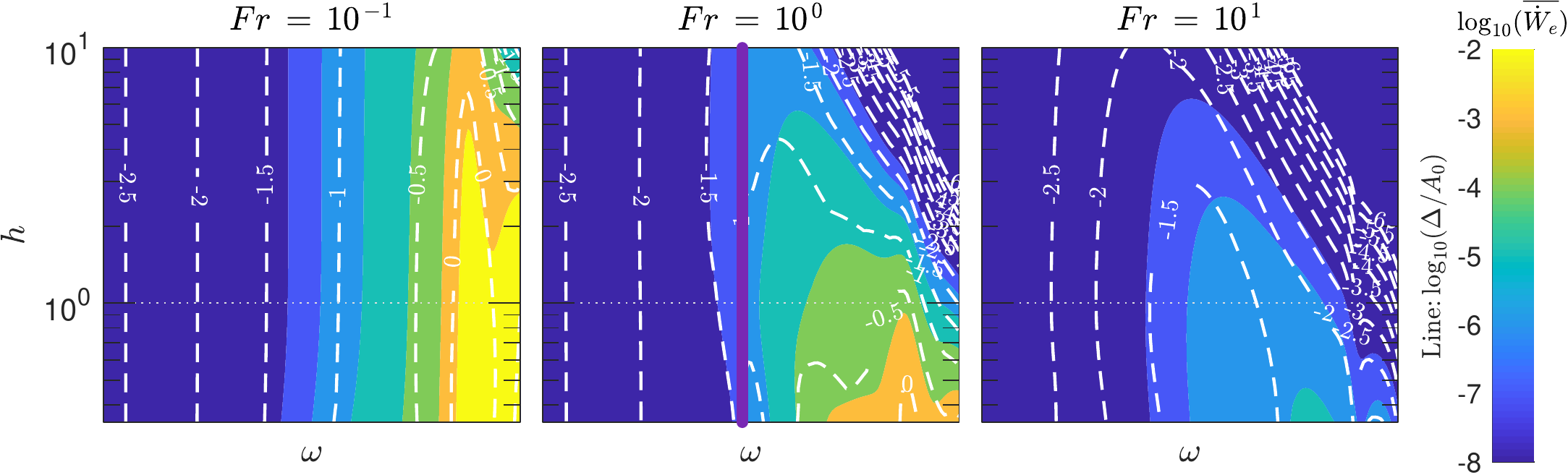}
		\put(-500,160){$(a)$}
        \put(-340,160){$(b)$}
        \put(-200,160){$(c)$}\\ \vspace{0.15cm}
		\includegraphics[width=\textwidth]{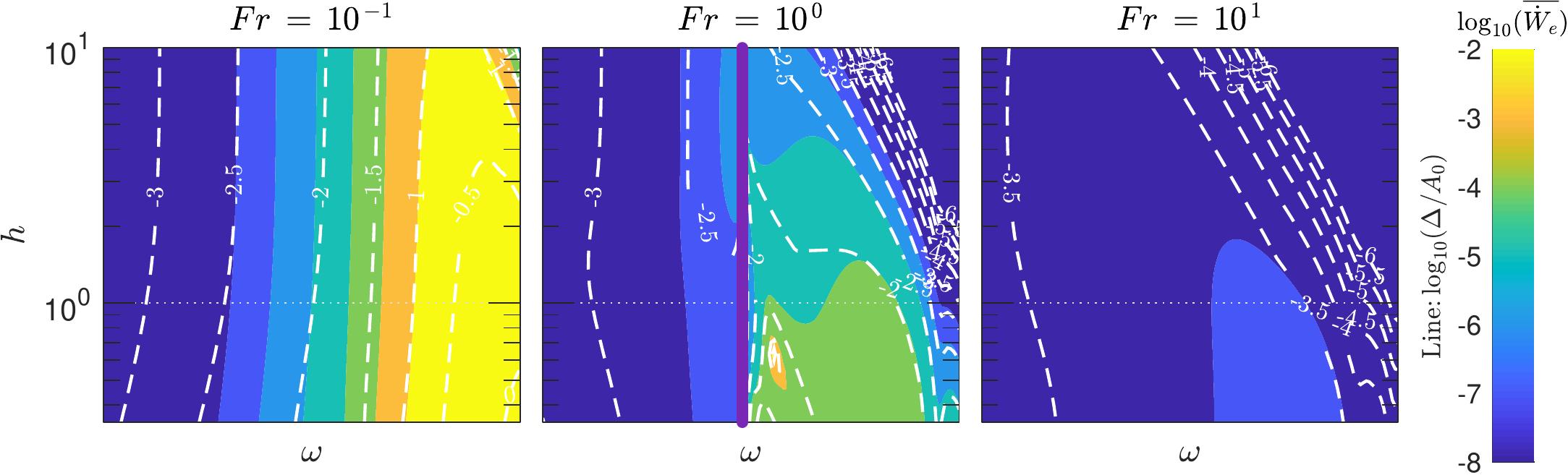}
		\put(-500,160){$(d)$}
        \put(-340,160){$(e)$}
        \put(-200,160){$(f)$}
        }
\caption{Dependency of the average electrical power output and maximum plate deflection on the submergence depth $h$ for different incoming wave frequencies $\omega$ and selected $Fr$. In subplots (a-c), $U^*=0.1$ and in (d-f) $U^*=10$, while the electrical parameters are fixed at the base values of $\alpha=0.5$, $\beta=1$, $\tau=0$.  
\label{Fig2_SLIDE27P}  }
\end{figure}
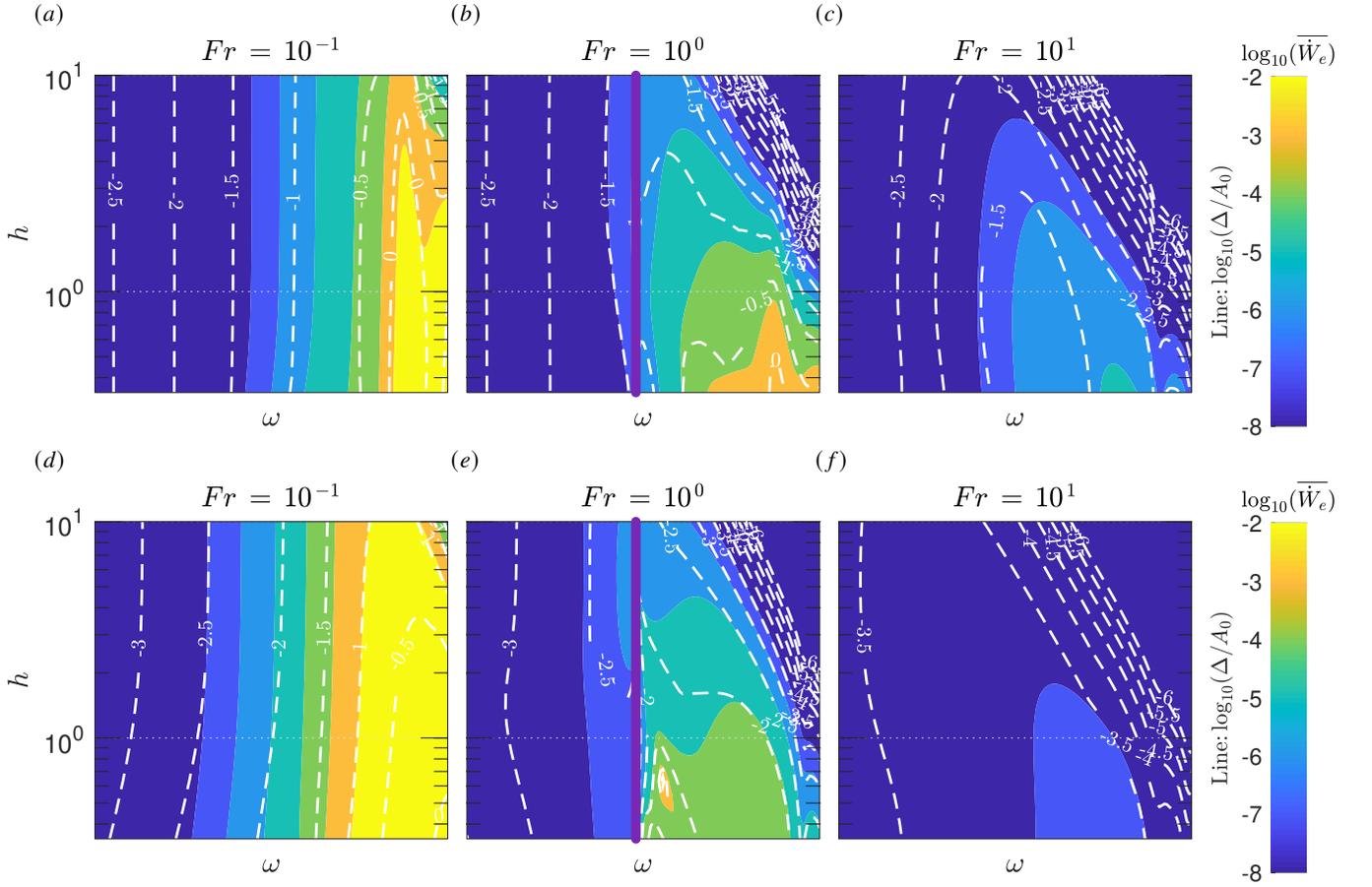

The plate's submergence depth changes its receptivity to the waves. Figure \ref{Fig2_SLIDE27P} shows the dependency of $\overline{\dot{W}_e}$ and $\Delta$ with the submergence depth and the wave frequency for subcritical, critical, and supercritical conditions. Other parameters are fixed at their reference values. For subcritical conditions, the results are almost independent of the submergence depth if $h<2$, equivalently when the submergence of the plate is less than its length. The same behavior is observed for all $U^*$ cases. The deviation is only for minimal submergence depth where the region with high $\overline{\dot{W}_e}$ extends to lower $\omega$ ranges. 
In contrast, a larger dependency on the submergence depth can be found for critical and supercritical flows. For these cases, the power output shows a higher decay with depth for less flexible plates than more flexible ones. This is due to a change in the energy harvesting mechanism from wave-induced fluttering motion in stiff plates to wave-triggering flapping motion in flexible plates. A flexible plate with flapping motion has a larger oscillation amplitude, and as a result, the combination of wave actuation and flow-induced flapping results in a higher amount of harvested energy despite the wave energy itself decaying with submergence depth. Yet, the benefit of the flapping motion declines fast with an increase of $Fr$, as the available wave energy weakens very fast with the depth. 

\begin{figure}
		\centering{
		\includegraphics[width=\textwidth]{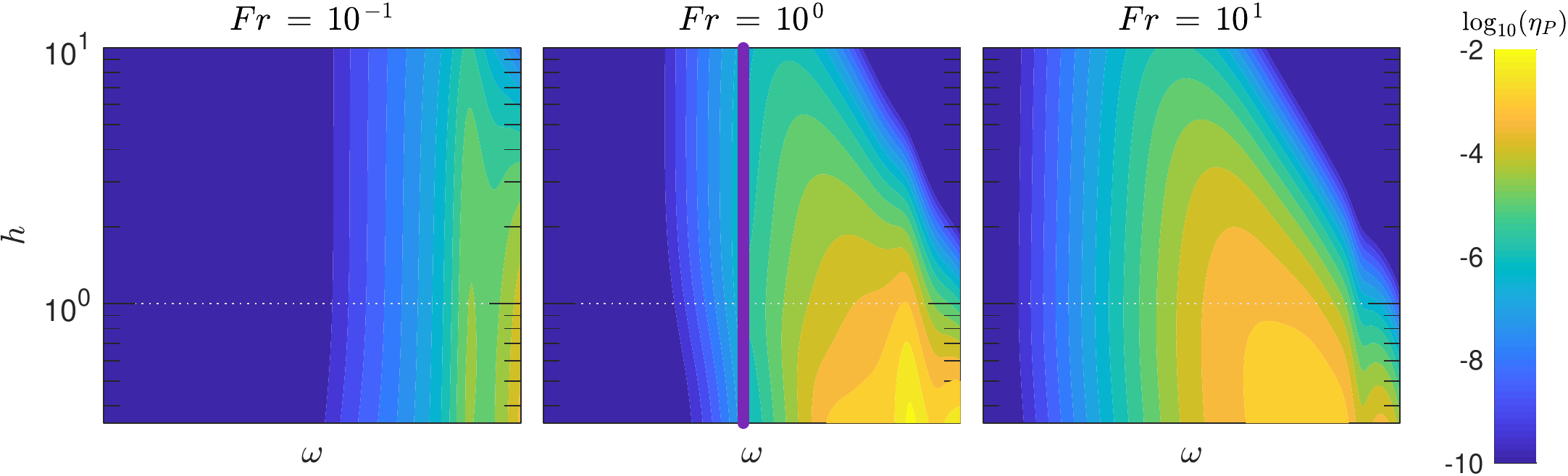}
						\put(-500,160){$(a)$}
        \put(-340,160){$(b)$}
        \put(-200,160){$(c)$}\\ \vspace{0.15cm}
		\includegraphics[width=\textwidth]{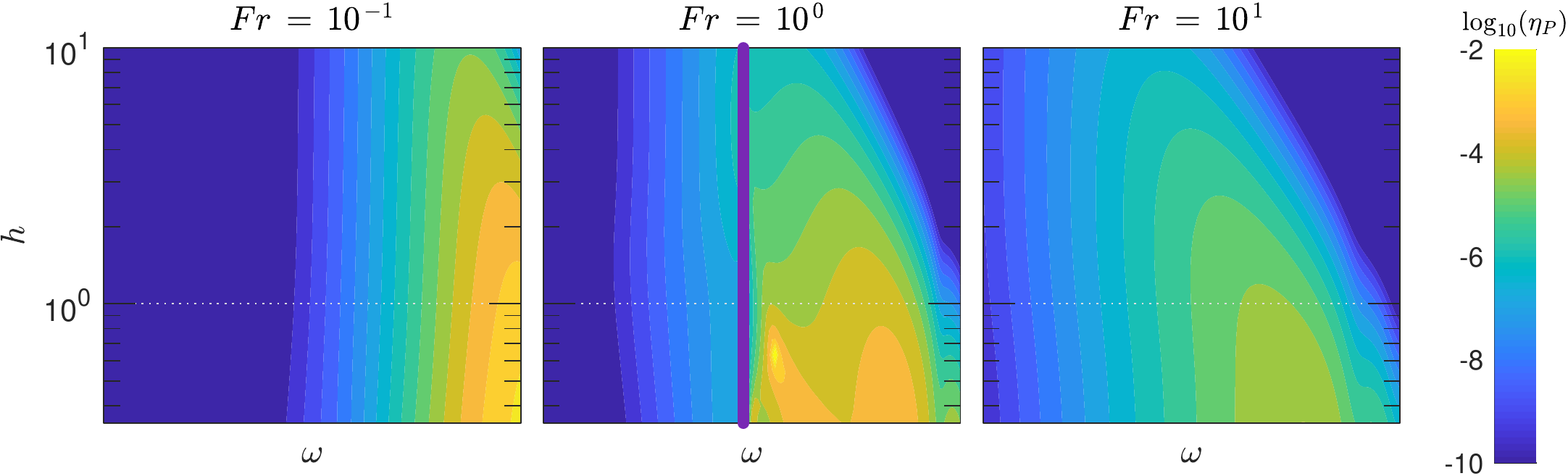}
		\put(-500,160){$(d)$}
        \put(-340,160){$(e)$}
        \put(-200,160){$(f)$}
		}
\caption{Similar to Fig. \ref{Fig2_SLIDE27P} expect it shows the contour of $\eta_P$.
\label{Fig2_SLIDE27E}  }
\end{figure}

The results suggest the flapping mode reaches the highest energy production if the plate is placed less than $h<1$ ( half of the plate's length) in subcritical and critical conditions. The energy extraction efficiency plots depicted in Figure \ref{Fig2_SLIDE27E} demonstrate that the highest energy efficiency throughout the depth occurs when the system operates under near-critical conditions and exhibits flapping mode vibration. The maximum attainable $\eta_P$ is smaller for stiffer plates and concentrates near the free surface for higher $Fr$ conditions. In practice, these optimal conditions can be achieved with longer plates  or faster incoming flow; in both cases, it is beneficial to place the plate near the free surface.

\subsection{Impact of electrical parameters}
\begin{figure}
		\centering{	\includegraphics[width=0.9\textwidth]{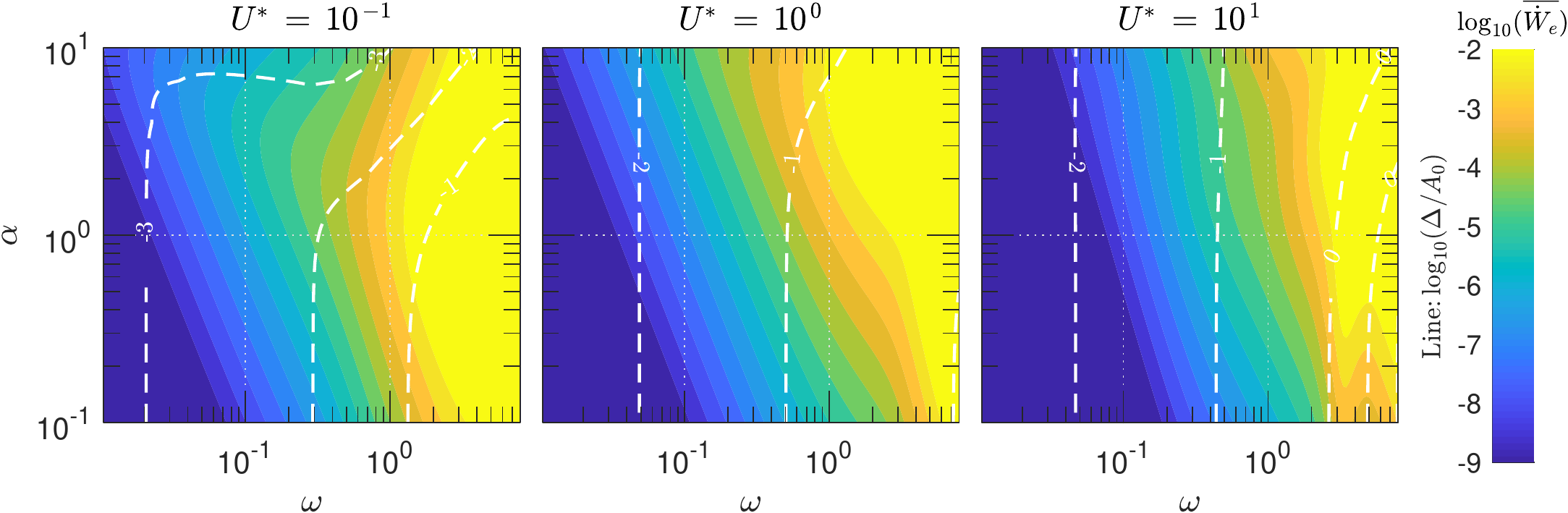}
				\put(-450,160){$(a)$}
        \put(-310,160){$(b)$}
        \put(-170,160){$(c)$}\\ \vspace{0.15cm}
		\includegraphics[width=0.9\textwidth]{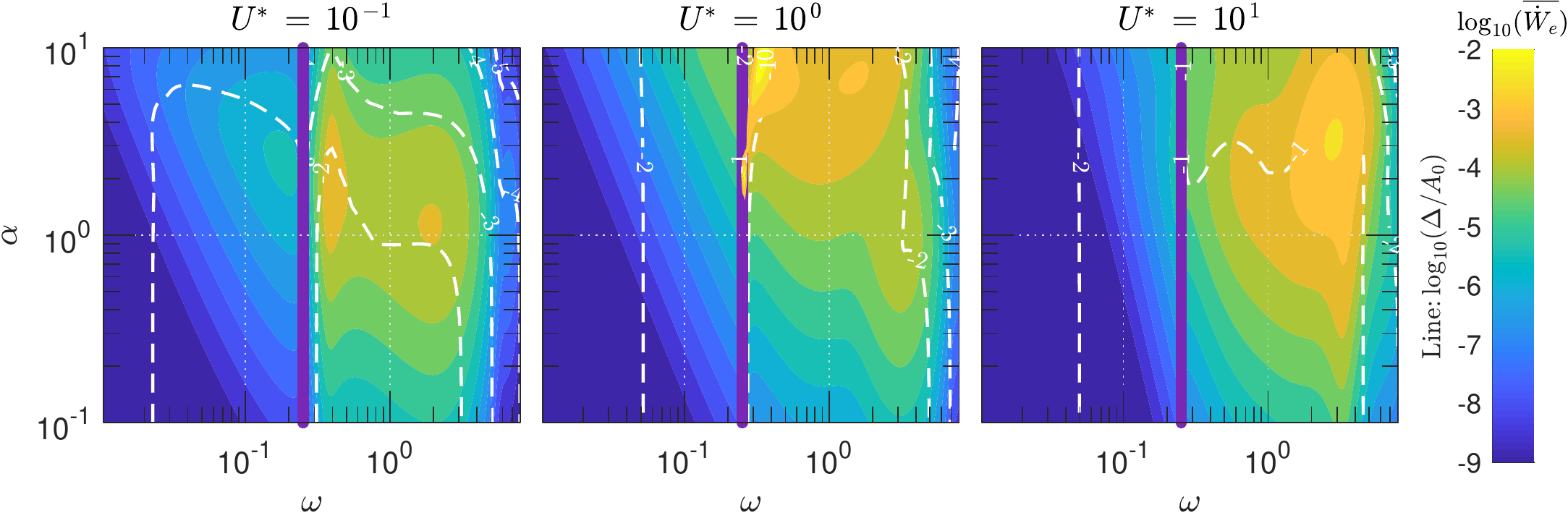}
\put(-450,160){$(d)$}
        \put(-310,160){$(e)$}
        \put(-170,160){$(f)$}
		}
\caption{The effect of electromechanical coupling coefficient $\alpha$ on the average power output and plate's  lateral deflection for three representative $U^*=0.1,1,10$ in the subcritical condition with $Fr=0.1$ (a-c) and critical condition with $Fr=1$ (d-e). $h=1.0$, $\beta=1$, and $\tau=0$. 
\label{Fig2_SLIDE28P}  }
\end{figure}

The coupling between a piezoelectric plate and a resonant circuit could substantially affect the energy harvesting performance of the piezoelectric flag \cite{xia2015fluid}. These effects are related to $\alpha$, the coupling coefficient, as well as $\beta$ and $\tau$, the resistive and inductive properties of the electrical circuit.
Figure \ref{Fig2_SLIDE28P} shows how electrical power output and the maximum plate deflection change with $\omega$ across a wide range of the electromechanical coupling parameter, $\alpha$. The other values are fixed at their nominal values. While $\alpha<0.5$ is expected in conventional piezoelectric materials, higher values can be achieved in multi-segmented systems such as M4, and Pelamis with conventional mechanical energy capturing derives. It is found that electrical power output is affected mainly by the wave frequency at low $Fr$ flow conditions and, to a much less degree, is dependent on $\alpha$ (Fig. \ref{Fig2_SLIDE28P}a-c). The central role of having larger $\alpha$ is to expand the energetic range to smaller $\omega$ while the energy level remains similar. The most energetic condition here is associated with stiffer plates and higher frequencies.

On the other hand, the electrical power output is highly dependent on $\alpha$ for higher $Fr$ cases when the plate is in the supercritical condition (right of the purple line in Fig.  \ref{Fig2_SLIDE28P}d-f). Here, the power output rapidly increases with $\alpha$ and attains its maximum values at $\alpha\approx 3$. However, the maximum energy output level is less than the value observed in the subcritical condition.   
The maximum deflection of the plate is almost independent of the $\alpha$ values except for certain conditions of large $\alpha$ and stiff plates (Fig.  \ref{Fig2_SLIDE28P}a) where the large electromechanical coupling results in higher effective bending stiffness and reduction of the wave-induced vibration amplitude. 

\begin{figure}
\centering {
\includegraphics[width=\textwidth]{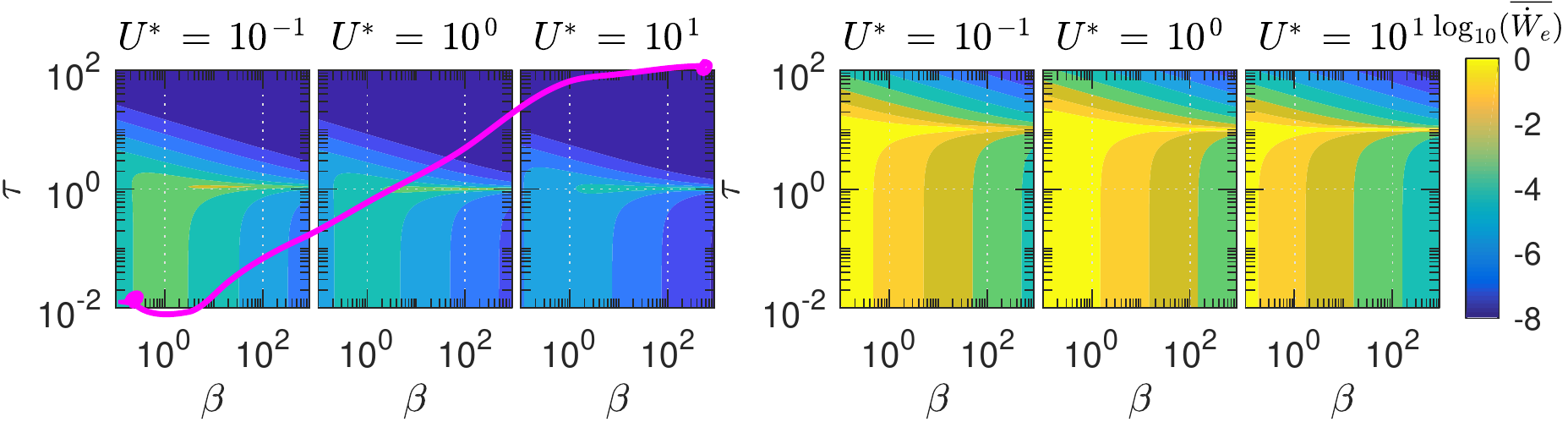}
\put(-500,130){$(a)$}
\put(-270,130){$(b)$}}

\caption{Dependency of the average electrical power output to inductive and resistive properties of the plate for $\omega=1$ (a) and 10 (b). Here $Fr=10^{-1}$, $h=1.0$ and $\alpha=0.5$ selected based on results in Fig. \ref{Fig2_SLIDE28P}.
\label{Fig2_SLIDE29PTAUBETA}  }
\end{figure}

The effects of resistive and inductive properties of the plate, quantified with $\beta$ and $\tau$, are shown in Fig. \ref{Fig2_SLIDE29PTAUBETA} for different $U^*$ values. In the following discussion, we
specifically focus on $Fr=0.1$, $h=1$ and  $\alpha=0.5$ to represent the energetic condition of Fig. \ref{Fig2_SLIDE28P}.  Two distinct $\omega$ values of 1 and 10 are tested where it is found that the electric inductance can enhance the output power over a narrow range of $\tau \omega \approx 1$. Over this range,  there is a destabilizing effect from the inductance, and as a result, the plate flutters with a higher amplitude. This trend is intact across different $U^*$, suggesting that the inductive properties can merely be adjusted based on the wave frequency, independent of the plate flexibility, to reach a better energy harvester. The very large $\tau$ limit acts similar to a short circuit and results in minimal effective coupling between electrical and mechanical fields. Consequently, the electric power output approaches zero regardless of $\beta$ for $\tau \gg \omega$. On the other hand, there is no inductance destabilization effect in small $\tau \ll \omega$ regimes and the electric circuit is purely resistive. In this case, the output power is just a  function of $\beta$  and reaches its maximum value when $\beta \omega \approx 1$.

\section{Conclusion}
This work explores the energy production from a two-dimensional electromechanical flexible plate placed in the proximity of a free surface and exposed to incident traveling gravity waves and incoming current. The  model represents a range of multi-segmented hybrid current/wave energy harvesting devices. In order to capture the interaction between the body, wake, and incident, diffracted, and radiated gravity waves, a Green function of a moving vortex near the free surface in inviscid flow is utilized. Through this theoretical model, we identified the optimal conditions for energy production and examined the predominant vibration modes of flexible plates situated near the free surface.  

Two distinct conditions with maximum energy production levels are identified: one is associated with the cantilever fluttering mode of the plate, and the other is very similar to the flow-induced flapping motion of the plate. The study demonstrates that the optimal energy production regimes exhibit similar trends and are influenced by changes in the Froude number. However, these energetic conditions occur at different non-dimensional flexibility values for the plate. The optimal flexibility for the cantilever fluttering mode remains relatively consistent across various incoming wave frequencies and Froude numbers. In contrast, the optimal flexibility for the flapping mode progressively decreases as the wave frequency increases. This divergent behavior between the two modes provides valuable insights about selecting and adjusting the mechanical parameters of hybrid wave/current energy harvesting devices.

The vibration amplitude and electrical power output are highly dependent on the Froude number. In particular, their contour plots show a distinct behavior in the subcritical, critical and supercritical flow conditions associated with $Fr<1, =1$ and $>1$, respectively. Moreover, the resultant surface waves from the plate presence in the flow, based on the $\omega Fr^2=\frac{1}{4}$ condition, determines the energy output level and efficiency of the piezoelectric plate. It is seen that the submergence depth should be smaller than the plate length to reach a good energy production level. The dependency on the submergence depth is more stringent for flow-induced vibration modes than the cantilever fluttering mode.  

The role of electrical parameters is more regular and does not substantially modify the optimal energy-producing conditions. Nonetheless, a larger electromechanical coupling coefficient modifies the dynamic characteristic of the system, especially more flexible plates vibrating in their flow-induced flapping mode. It mainly broadens the frequency range corresponding to near-maximum energy production behavior. The optimal resistive value of the electric network related to wave frequency $\beta {\omega} \approx 1$ and the optimal electric inductance is associated with near resonance condition in the electric network, namely when $\tau \omega \approx 1$. 

The results provided in this paper could be helpful in the estimation of the operating threshold of hybrid wave/current energy harvesting devices that resemble a piezoelectric plate.  It can also be employed to adjust the system's parameters, in particular the length of the device, to reach one of the most energy-producing conditions. The current theoretical prediction using the inviscid flow theory and linear approximation should be tested with future experiments and fully nonlinear fluid-structure interaction simulations \citep{vahab2021fluid}. In addition, the proposed model and the dispersion relation are insufficient to explore the energy production of the system at the critical condition of $\omega Fr^2=\frac{1}{4}$ where a group of reflected and radiated waves combines into one wave with a zero group velocity and no ability to transport wave energy. It would be an interesting extension of this study to theoretically explore energy production under this condition. 

{The predictions from the current linear model suggest that with current piezoelectric materials, the energy efficiency of the device is less than $5\%$, much smaller than other conventional hydrokinetic energy harvesters \citep{drew2009review}. It is observed that with better-engineered electromechanical converters, and larger $\alpha$ coefficient and with the use of the inductive elements in the electrical conversion unit, the energy efficiency of the device can be much improved and becomes close to $40\%$. The flexible piezoelectric plates can also be employed as unsteady thrust-producing devices if they are attached to the side of a boat or ship and undergo the combined plunge and pitch motions \citep{grue_propulsion_1988, belibassakis2013hydrodynamic}. In addition, these types of energy converters have a unique property that they can easily be deployed or collected depending on the sea condition.}  The current model can be employed to determine the performance of hybrid wave/current energy harvesters as both the thrust-generating and energy-producing devices in these conditions.
The three-dimensional effects of finite-width plates could also modify wave diffraction and energy production. Including the three-dimensional effects into the current framework is the subject of ongoing research.

\appendix
\section{}\label{appA}
{In the appendix, we assume all variables are dimensional and  we derive the most general equation of the system assuming inviscid irrotational two-dimensional flow with $y=0$ placed at the air-sea interface pointing upward. The velocity $\bm{u}$ is  expressed based on the
the gradient of a scalar potential $\Phi$,
as $\bm{u}\,=\,\nabla \Phi$
and the conservation of mass requires that the potential satisfies Laplace’s equation
\begin{align} 
\nabla^2\Phi\,=\,0
\label{A2}
\end{align}  

The pressure can be calculated from, 
\begin{align} 
-\frac{P}{\rho}\,=\,g y+\frac{\partial \Phi}{\p t}+\frac{1}{2} |\nabla\Phi|^2+B(t)
\label{A3}
\end{align}  
where $B(t)$ is an arbitrary function of $t$ that can be emitted by redefinition $\Phi$ without affecting the velocity field.

We have two types of boundaries: the air–water interface, which is also
 called the free surface, and the wetted surface of an impenetrable solid which in this study is the surface of a thin piezoelectric plate. Along both surfaces, the flow can have only relative motion in the tangential direction and the instantaneous equation of  boundary conditions can be written as 
\begin{align} 
F(\bm{X}(t),t)=y-\xi(x,t)\,=\,0
\label{A4}
\end{align}  
where $\xi$ is the vertical coordinate measured from $y=0$, and $\bm{X}(t)$ is
a point on the interfaces moving with velocity $\bm{U}$. By taking the derivative of Eq. \ref{A4} and with the assumption of only tangential relative motion at the interfaces (equivalently $\bm{U}\cdot\nabla F\,=\,\bm{u}\cdot\nabla F$) we can  write the condition of moving surfaces as 
\begin{align} 
\frac{\p \xi}{\p t}\,+\,\pard{\Phi}{x}\pard{\xi}{x}\,=\, \pard{\Phi}{y}
\label{A5}
\end{align}

Eq. \ref{A5} defines the kinematic boundary conditions on the plate surface and also on the free surface.   On both boundaries, it is necessary to add the dynamic boundary condition. On the free surface, with atmospheric pressure $P_a$, we can write
\begin{align} 
-\frac{P_a}{\rho}\,=\,g\eta +\frac{\partial \Phi}{\p t}+\frac{1}{2} |\nabla\Phi|^2\hspace{1cm}\text{on}\,\,\,\,\,\,y\,=\,\eta
\label{A6}
\end{align}
where $\eta$ is the surface wave height. This equation can be combined with Eq. \ref{A5} and with the assumption of $P_a=$ constant, we can write the dynamic condition on the free surface as,
\begin{align} 
\pardo{\Phi}{t}
\,+\,
g \pard{\Phi}{t}
\,+\,
\pard{\,}{t}\bm{u}^2\,+\,\frac{1}{2}\bm{u}\cdot\nabla \bm{u}^2\,=\,0\hspace{1cm}\text{on}\,\,\,\,\,\,y\,=\,\eta
\label{A7}
\end{align}

With the presence of a current with velocity $u_c$ along the $x$ axis, we can use the superposition technique and define $\Phi=\Phi_w+u_c x$, where $\Phi_w$ is the velocity potential due to the surface waves and $u_c x$ captures the effect of the current. 
To linearize the problem, we assume that certain physical scales of motion can be anticipated \emph{a priori} and define the following non-dimensional  quantities,
\begin{align}    
\{\hat{x},\hat{y}\}\,=\,\frac{2 \pi \{x,y\}}{\lambda}, \hspace{0.5cm} 
\hat{t}\,=\,\omega t, \hspace{.5cm} 
\hat{\eta}\,=\,\frac{\eta}{A}, \hspace{.5cm} 
\hat{\Phi}\,=\, \frac{2 \pi \Phi}{A\omega \lambda} 
\label{A8}
\end{align}
where $\lambda$, $\omega$, and $A$ are the typical values of wavelength, frequency, and free-surface amplitude, respectively. We can rewrite Eqs. \ref{A2}, \ref{A4} and \ref{A6} based on $\hat{\Phi}_w$ as,

\begin{align} 
\nabla^2\hat{\Phi}_w\,&=\,0,\\
\frac{\p \hat{\xi}}{\p \hat{t}}\,+\,u_c \pard{\hat{\eta}}{\hat{x}}\,&+\,\epsilon \left(\pard{\hat{\Phi}_w}{\hat{x}}\pard{\hat{\eta}}{\hat{x}} \right) \, =\, \pard{\hat{\Phi}_w} {\hat{y}} \hspace{1cm}\text{on}\,\,\,\,\,\,\hat{y}\,=\,\epsilon\hat{\eta} \\
\frac{\partial \hat{\Phi}_w}{\p \hat{t}}
+\left(\frac{2\pi g}{\omega^2\lambda}\right) \hat{\eta} \,&+\,
\frac{\epsilon}{2} (\hat{\nabla}\hat{\Phi}_w)^2\,=\, -\frac{2\pi P_a}{\rho A\omega^2 \lambda} \hspace{1cm}\text{on}\,\,\,\,\,\,\hat{y}\,=\,\epsilon\hat{\eta}
\label{A9}
\end{align}
where $\epsilon=2\pi A/\lambda$ = wave slope. If we assume $\epsilon \tt 1$, we can drive the linear equations to $O(\epsilon)$, which in their non-dimensional forms based on $u_c$ and $b$ will be similar to Eqs. \ref{eq:POT}, \ref{eq:KBC} and \ref{eq:DBC}. }

\noindent{\bf Acknowledgements:} {The authors would like to acknowledge the Florida State University Research Computing Center and NSF ACCESS computational resources through grant number CTS200043 on which these simulations were carried out.}

\noindent{\bf Funding:} {This work is partially supported by NSF grant CBET-1943810.}

\end{document}